\documentclass[prd,preprint,amsmath,nofootinbib,superscriptaddress]{revtex4-1}
\pdfoutput=1
\usepackage{bm}
\usepackage{slashed}
\usepackage{amsmath,latexsym}
\usepackage{multirow}
\usepackage{graphicx}
\usepackage{hyperref}
\usepackage[usenames,dvipsnames]{color}

\def\OMIT#1{}

\newcommand{\bea}{\begin{eqnarray}}
\newcommand{\eea}{\end{eqnarray}}

\newcommand{\beq}{\begin{equation}}
\newcommand{\eeq}{\end{equation}}
\newcommand{\mo}{\mathcal{O}}
\newcommand{\nn}{\nonumber} 

\newcommand{\vacproj}{|0\rangle \langle 0 |}

\begin{document}

\title{Semi-Inclusive Wino and Higgsino Annihilation to LL$^\prime$}

\author{Matthew Baumgart}
\affiliation{Department of Physics, Carnegie Mellon University,
    Pittsburgh, PA 15213}
\affiliation{New High Energy Theory Center, Rutgers University, Piscataway, NJ 08854}
\author{Varun Vaidya \vspace{0.4cm}}
\affiliation{Department of Physics, Carnegie Mellon University, Pittsburgh, PA 15213}


\begin{abstract}
We systematically compute the annihilation rate for winos and higgsinos into the final state relevant for indirect detection experiments, $\gamma + X$.  The radiative corrections to this process receive enhancement from the large Bloch-Nordsieck-Violating Sudakov logarithm, $\log(2 M_\chi/M_W)$.  We resum the double logs and include single logs to fixed order using a formalism that combines nonrelativistic and soft-collinear effective field theories.  For the wino case, we update an earlier exclusion adapting results of the HESS experiment.  At the thermal relic mass of 3 TeV, LL$^\prime$ corrections result in a $\sim$30\% reduction in rate relative to LL.  Nonetheless, single logs do not save the wino, and it is still excluded by an order of magnitude.  Experimental cuts produce an endpoint region which, our results show, significantly effects the higgsino rate at its thermal-relic mass near 1 TeV and is deserving of further study.
\end{abstract}

\maketitle

\section{Introduction}

While the Standard Model (SM) lacks a viable dark matter (DM) candidate, it is straightforward to extend it to a theory that does.  The fact that a weak-scale particle charged under the electroweak group can simply freeze out of the primordial plasma with the correct relic abundance has been dubbed the ``WIMP Miracle'' ({\it cf.}~\cite{Dimopoulos:1990gf}).  Since the dynamics that set the electroweak scale are presently unknown, it is natural to go a step further and propose that WIMP (Weakly-Interacting Massive Particle) dark matter arises from the sector that solves the hierarchy problem.  Supersymmetry (SUSY) is the classic example of this `two-birds with one stone' approach to beyond the Standard Model physics.  However, the absence of superpartner discovery at the LHC and elsewhere has brought the public perception of the theory into some disrepute.  One may counter though, that this very notion of SUSY's having a successful, simple DM candidate predicts precisely the current experimental situation.  The dark matter is necessarily the lightest supersymmetric particle (LSP), and its most straightforward realization is the neutralino.  In the absence of large mixing, we can discuss each of the three neutralino states as its own case.  The bino overcloses the universe in the absence of nearly-degenerate sfermions.  This leaves wino and higgsino, the two cases we consider here.  If we want them to be the advertised thermal relic, in the absence of degeneracies in the SUSY spectrum, this fixes their masses to be $M_{\rm wino} \equiv (M_{\chi})$ = 2.7-2.9 TeV or $M_{\rm higgsino} \equiv (M_{\chi})$ = 1 TeV \cite{Fan:2013faa,constraint}.  Thus, the simple SUSY dark matter story has an electroweak LSP at or above 1 TeV, and thus has no observable collider signature for the foreseeable future.

While TeV-scale, weakly-interacting states are out of range of the current generation of colliders, they represent a big target of opportunity for indirect detection experiments.  Observation of a nearly monochromatic line of photons at these energies would be a smoking gun of new physics.  The possibility of such experiments to set stringent limits on the wino scenario was explored in \cite{Fan:2013faa,Cohen:2013ama}.  By combining a numerical calculation of the Sommerfeld enhancement with a tree-level annihilation of WIMPs to $\gamma \gamma + \frac{1}{2} \gamma Z$, they found that the HESS experiment ruled out the wino by a factor of 15 under the assumption of an NFW dark matter halo profile \cite{hess}.  In particular though, \cite{Cohen:2013ama} extended their annihilation calculation to one-loop and found a reduction in rate by $\sim 4\times$ at the thermal relic mass of 3 TeV.  In order to test such a large radiative correction, in our previous papers (\cite{Baumgart:2014vma,Baumgart:2014saa}), we developed an effective field theory (EFT) approach to resum the large Sudakov double-logs that appear, $\log(2 M_\chi/M_W)$, factorizing their contribution from the Sommerfeld factors.  Although the gauge bosons in the theory are electroweak, the approach is based on two different EFTs developed for QCD, non-relativistic QCD (NRQCD) \cite{NRQCD} and soft-collinear effective theory (SCET) \cite{SCET}.  Since the experiment only detects one of the hard photons from annihilation and its resolution is on the order of a few hundred GeV, we computed the semi-inclusive annihilation rate, $\chi^0 \chi^0 \rightarrow \gamma + X$.  We found the overall correction from resummed double logs to be modest, with a few percent reduction in the cross section at the relic density mass of ~3 TeV relative to tree level (+ Sommerfeld enhancement).  Thus, the factor of $\sim15$ exclusion by HESS remained, unless one invoked a profile with coring $\gtrsim$ 1.5 kpc, in tension with recent simulations \cite{DiCintio:2013qxa,Kuhlen:2012qw,Marinacci:2013mha}.

In between the release of our two papers, two other groups produced calculations using effective field theory for wino annihilation \cite{Bauer:2014ula,Ovanesyan:2014fwa} to $\gamma \gamma + \frac{1}{2} \gamma Z$.\footnote{Ref.~\cite{Bauer:2014ula} actually treated an SU(2) scalar triplet, but as we are in the nonrelativistic limit and only annihilating a single spin channel in the fermion case, the calculation is identical.}  Both papers reported a much larger reduction, getting roughly half the rate found at tree level.  This need not be a contradiction as our semi-inclusive cross section necessarily includes more processes.  Nonetheless, to improve our earlier, leading log (LL) resummed result, we now compute the single-log contributions.  Despite being formally subleading in the power counting parameter, working at single-log order brings in new effects which have the potential to be large: 1) the possibility of the photon to fragment into pairs of SM particles, 2) accounting for real-emission processes that make $E_\gamma$ sufficiently less than $M_\chi$ that events wind up outside the signal bin.  We will find, in fact, that 2) is numerically important.  While this is not enough to prevent the wino from being ruled out by HESS by an order of magnitude, it does prompt us to undertake a future study to calculate endpoint logarithms which are different from the $\log(2 M_\chi/M_W)$ contributions we have been including and resumming thus far \cite{futuremv}.  We also extend our formalism to calculate the semi-inclusive rate of higgsino annihilation.  Since the thermal relic mass is just 1 TeV, the log expansion is less accurate and we will see that single logs introduce a sizable correction.

In Section \ref{sec:eft}, we briefly review the EFT for WIMP annihilation to $\gamma + X$, and present our previous results for the wino with LL resummation along with single-log contributions at fixed order, providing an LL$^\prime$ result. Section \ref{sec:hino} extends this to the higgsino.  In Section \ref{sec:conc}, we give the plots of the cross section as a function of the WIMP mass. While the splitting between charged and neutral wino states are fairly model independently around 170 MeV, we give plots for the higgsino-case in two different limits of mass splittings. We also discuss the importance of corrections arising out of finite detector resolution and ways that these can be handled in a controlled manner.  In the Appendix, we collect some technical results on the neutralino mass matrix in the MSSM, Sommerfeld enhancement, and our treatment of the photon fragmentation down to zero invariant mass.

\section{Effective Field Theory of Heavy WIMP annihilation}
\label{sec:eft}

We lay out the details of our effective field theory approach and factorization theorem in \cite{Baumgart:2014vma,Baumgart:2014saa}.  The EFT is a hybrid of NRQCD and SCET.\footnote{The SCET we use is SCET-II, as soft and collinear fields have the same virtuality.  This is because $M_W$, which sets the collinear scale, also provides an IR cutoff, beyond which soft $W$s cannot fall.}  The former is necessary to handle the heavy, slowly-moving initial state and is organized as a power counting in $v$, the WIMP velocity.  The latter resums the large kinematic logarithms that arise from our highly-boosted final state particles, with a power-counting in $\lambda = M_W/M_\chi$.  One immediate benefit of this approach is that it disentangles the different physical effects arising from the same fields in the underlying theory.  Since we are quantifying the nonperturbative physics of both Sommerfeld enhancement and Sudakov resummation and since loops of $W$ boson contribute to both, the EFT makes it manifest which nontrivial process a given diagram is contributing to.  In Table \ref{tab:table1}, we list the effective theory fields and their power counting.\\

\begin{table}[h!]
  \begin{center}
    \caption{Momentum scalings for the EFT fields. The SCET field momenta are in light cone-coordinates.  The collinear gauge boson and WIMP fields appear explicitly in our operators, so we have included our notation for them below.  Soft gauge bosons, $n \cdot A_n$, enter our calculation through soft Wilson lines.  Since we solve for the Sommerfeld enhancement with an instantaneous potential, this accounts for the action of the potential modes.}
   \label{tab:table1}
\vspace{0.2in}   
\begin{tabular}{cc|c|}
\cline{1-3}
\multicolumn{1}{ |c }{\multirow{2}{*}{SCET} } &
\multicolumn{1}{ |c| }{$B_n^\perp$: Collinear} & $S_n \supset$ Soft   
\\ \cline{2-3}
\multicolumn{1}{ |c  }{($p^+$, $p^-$, $p_{\perp}$)}                        &
\multicolumn{1}{ |c| }{$M_{\chi}$(1,$\lambda^2$,$\lambda$)} & $M_{\chi}(\lambda,\lambda,\lambda)$ 
\\ \cline{1-3}
\multicolumn{1}{|c}{\multirow{2}{*}{NRQCD} } &
\multicolumn{1}{ |c| }{Potential} & $\chi$: WIMP
\\ \cline{2-3}
\multicolumn{1}{ |c  }{(E, p)}                        &
\multicolumn{1}{ |c| }{ ($M_{\chi} v^2, M_{\chi}v$)} & { ($M_{\chi} v^2, M_{\chi}v$)}
\\ \cline{1-3}
 
\end{tabular}
\end{center}
\end{table}

\begin{figure}
\centerline{\scalebox{.5}{\includegraphics{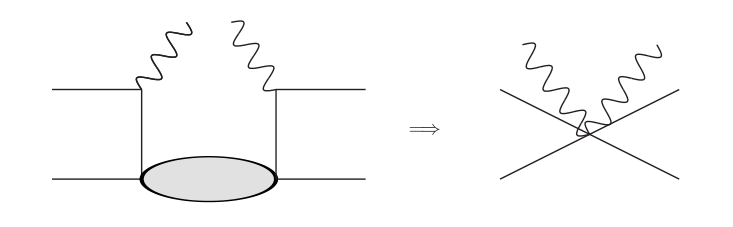}}}
\vskip-0.2cm
\caption[1]{Integrating out the recoil jet. The open curly lines correspond to the observed photon which is accompanied 
by any amount of soft or collinear radiation.}
\label{semi} 
\end{figure}

Additionally, we note that for the experiments under consideration, resolution of the photon signal is sufficiently poor that we cannot distinguish between strict two-body ($\gamma \gamma$ or $\gamma$Z) annihilation and processes where the photon either recoils against or emits soft or collinear $W$s.  For this reason, we make use of the Operator Product Expansion (OPE) to calculate the semi-inclusive annihilation rate $\sigma(\chi^0 \chi^0) \rightarrow \gamma + X$ ({\it cf.}~Fig.~\ref{semi}).   
The recoil state will not lead to any IR divergences dependent on $M_W$, even at the single log level.  Since we are totally inclusive in the recoil- (or $\bar n$-) sector, and since the initial state has no $\bar n$ particles, by taking the OPE, we are calculating a recoil-sector vacuum expectation value.  Thus, the only operator in the expansion that can contribute to the vev is proportional to the identity and is therefore IR safe.\footnote{We thank Ira Rothstein for this argument \cite{book_ira}.}  We can perform the OPE by matching to the semi-inclusive rate to produce a photon in the full theory at the scale $2M_\chi$.  The procedure is identical in both the wino and higgsino cases, but the different representations of these states, along with the higgsino's nonzero hypercharge, give us a different basis of operators.  We now discuss both cases. 

\subsection{Wino Operators}
\label{subsec:winolike}

The first case we consider is that of a wino-like neutralino.  This covers the possibility of dark matter arising as the LSP in the MSSM (or an extension) as long as it is sufficiently split from other states that we can ignore the effects of coannihilation and mixing.  Operationally, though, we are simply adding an SU(2) triplet to the SM, and so this scenario could arise as a more minimal model of DM.  We will nonetheless refer to this state as the wino.  The details of the operator basis and factorization theorem along with the analysis of the cross-section at leading log were presented in \cite{Baumgart:2014vma},\cite{Baumgart:2014saa}. Here we briefly review this case and extend the analysis to LL$^\prime$, including the single-log contributions as fixed order corrections to the resummed double log.  The operator basis dressed with Wilson lines is 
\bea
O_1&=& \left( \bar  \chi \gamma^5 \chi \right) \vacproj \left( \bar \chi \gamma^5 \chi \right) B^A B^ A\nn\\
O_2&=& \frac{1}{2}\Big\{(\bar \chi \gamma^5\chi) \vacproj  (\bar \chi_{A^\prime} \gamma^5 \chi_{B^\prime}) + (\bar \chi_{A^\prime} \gamma^5 \chi_{B^\prime}) \vacproj (\bar \chi \gamma^5\chi) \Big\} B^{\tilde A} B^{\tilde B} \nn\\
&&S_{v A^\prime A}^\top \, S_{vB B^\prime} \, S_{n  \tilde A A}^\top \, S_{n B \tilde B}\nn\\
O_3&=& \left( \bar \chi_C\gamma^5 \chi_D \right) \vacproj \left( \bar \chi_D \gamma^5 \chi_C \right) B^A B^A \nn\\
O_4&=& (\bar \chi_{A^\prime} \gamma^5 \chi_C) \vacproj  (\bar \chi_C \gamma^5 \chi_{B^\prime})  B^{\tilde A} B^{\tilde B} \, S_{v A^\prime A}^\top \, S_{vB B^\prime} \, S_{n  \tilde A A}^\top \, S_{n B \tilde B}. 
\label{eq:winoops}
\eea
The vacuum insertion approximation enforces that the WIMP fields annihilate the initial state and holds up to corrections of $\mo(v^2)$.  In general, we will drop the vacuum projector.  There is also an implicit Lorentz contraction and projection of the final state onto a single-photon state in the $n$ direction,
\beq
 B_{n\,\mu}^{\perp A} B^{\perp \mu \, B}_n \equiv \sum_X  B_{n\,\mu}^{\perp A} \mid \gamma+X\rangle\langle \gamma+X\mid B^{\perp \mu \, B }_n,
 \label{eq:photproj}
\eeq
and without ambiguity, we drop the $n$ and $\perp$ from $B^{\perp A}_n$.
In preparation for computing the anomalous dimensions, it is useful to decompose the operators in Eq.~\ref{eq:winoops} into the soft and collinear sectors,
\bea
O_s^a&=&S_{v A^\prime A}^T S_{vB B^\prime}S_{n  \tilde A A}^TS_{n B \tilde B}, \ \ \ \ O_s^b=\delta_{\tilde A \tilde B} \delta_{A^\prime B^\prime}
\nn \\
O_c^a &=& B_{\tilde A} B_{\tilde B}\nn\\
O_c^b&=& B_D \, B_D \, \delta_{\tilde A \tilde B}.
\label{eq:wscops}
\eea
The cross-section in terms of this basis is 
\bea
&& \frac{1}{E_\gamma}  \frac{d\sigma}{dE_\gamma} = \frac{1}{4M_\chi^2 v} \langle 0 | O^a_s | 0 \rangle \Bigg[ \int d n \cdot p \, \Bigg\{ C_2(M_\chi, n \cdot p) \langle p_1 p_2 \mid \frac{1}{2}\Big\{(\bar \chi \gamma^5\chi) \, (\bar \chi_{A^\prime} \gamma^5 \chi_{B^\prime}) \nn \\
&+& (\bar \chi_{A^\prime} \gamma^5 \chi_{B^\prime}) (\bar \chi \gamma^5\chi) \Big\}(0) \mid p_1 p_2 \rangle + C_4(M_\chi, n \cdot p) \langle p_1 p_2 \mid (\bar \chi_{A^\prime} \gamma^5 \chi_C) \, (\bar \chi_C \gamma^5 \chi_{B^\prime}) (0) \mid p_1 p_2 \rangle \Bigg\} F^\gamma_{\tilde A \tilde B}\left( \frac{2E_\gamma}{n \cdot p} \right) \Bigg] \nn\\
&&+  \Bigg[ \int d n \cdot p \,\Bigg\{ C_1(M_\chi, n \cdot p)\langle p_1 p_2 \mid (\bar \chi \gamma^5 \chi) \, (\bar \chi \gamma^5 \chi) (0) \mid p_1 p_2 \rangle +C_3(M_\chi, n \cdot p) \nn \\
&\times& \langle p_1 p_2 \mid (\bar \chi_C \gamma^5 \chi_D) \, (\bar\chi_D \gamma^5 \chi_C) (0) \mid p_1 p_2 \rangle\Bigg\} F_\gamma\left( \frac{2E_\gamma}{n \cdot p} \right)  \Bigg],
\label{crosswino}
\eea
where $F^\gamma_{\tilde A \tilde B}$ is a fragmentation function defined by
\bea
F^\gamma_{\tilde A \tilde B}\left( \frac{n\cdot k}{n\cdot p} \right) &=& \int \frac{ dx_-}{2\pi}e^{in \cdot p \, x_-}\langle 0 \mid B^{\perp \mu}_{\tilde A}(x_-) \mid \gamma(k_n)+X_n \rangle \nn \\
&\times& \langle \gamma(k_n)+X_n \mid  B^\perp _{\mu \tilde B}(0)  \mid 0 \rangle ,
\label{eq:fragdefw}
\eea 
where $n \cdot p$ is the total +-momentum flowing in the $n$-direction and $F_\gamma = F^\gamma_{\tilde A \tilde B} \delta_{\tilde A \tilde B}$.  The $C_i$ are just the Wilson coefficients of the full operators given in Eq.~\ref{eq:winoops}.  As we will detail below, there is a simple relation between them at tree level that allows us to determine them at that order by a single calculation.

\subsection{Wino Anomalous Dimensions}

The advantage of working with the basis in Eq.~\ref{eq:wscops} becomes apparent when we work at loop level.  We can consider the soft and collinear sectors separately.
However, soft and collinear modes have the same virtuality and hence the divergences that arise from the factorization of the soft sector from the collinear cannot be regulated by dimensional regularization, which respects boost symmetry. Hence, we need to introduce a rapidity regulator, which manifestly breaks boosts \cite{RRG}. This requires a corresponding factorization scale which we call $\nu$.  As it arises from our artificial distinction between soft and collinear modes, it will necessarily drop out of the anomalous dimensions of the full operators in Eq.~\ref{eq:winoops}, after we recombine the separate soft and collinear operators of Eq.~\ref{eq:wscops} to make them.  This provides a useful cross check on the calculation.

\begin{figure}
\centerline{\scalebox{0.6}{\includegraphics{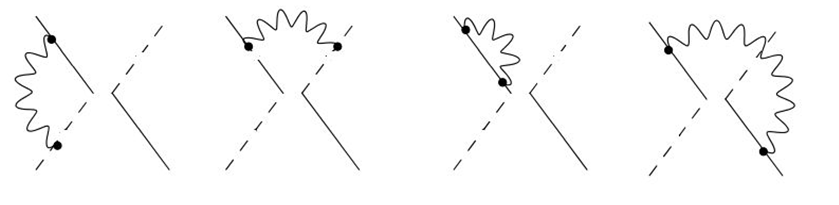}}}
\vskip-0.2cm
\caption[1]{The diagrams contributing to $O_s^a$. The solid/dashed lines indicate time/light-like Wilson lines.}
\label{softa} 
\end{figure}

We now calculate the matrix elements of the operators defined above, including the single log contributions.  The diagrams contributing to the operator $O_s^a$ are shown in Fig. \ref{softa}.
\bea
\langle 0|O_s^a|0 \rangle  &=&\!  \delta_{ A^\prime \tilde A} \delta_{B^\prime \tilde B} \!+ \{\delta_{\tilde A \tilde B} \delta_{A^\prime B^\prime} \!-\! 3\, \delta_{ A^\prime \tilde A} \delta_{B^\prime \tilde B}\}\frac{\alpha_W}{\pi}\left[2 \log(\frac{\nu}{\mu}) \log(\frac{\mu}{M_W})\!+\log^2(\frac{\mu}{M_W})-\log(\frac{\mu}{M_W})\right]\nn\\
\eea
\begin{figure}
\centerline{\scalebox{0.6}{\includegraphics{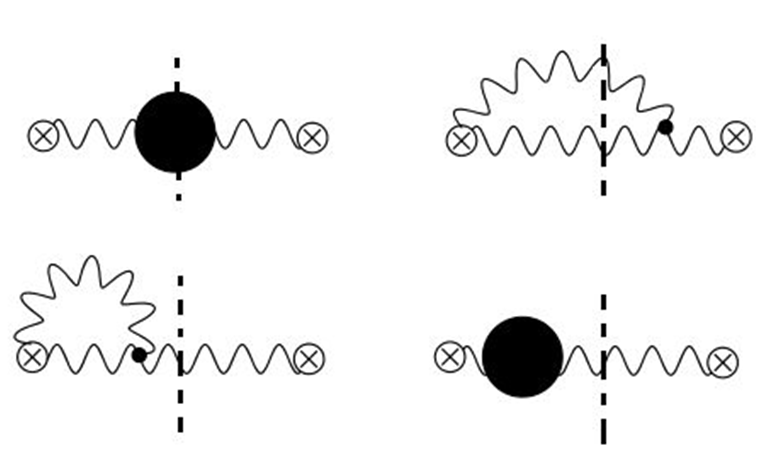}}}
\vskip-0.2cm
\caption[1]{The diagrams contributing to $O_c^a$. The dark blob contains fermion, scalar and gauge boson loops. }
\label{colla} 
\end{figure}
Thus, we radiatively generate the singlet color structure which demonstrates mixing between the singlet and nonsinglet operators in the soft sector.

The diagrams for the operator $O_c^a$ are shown in Fig.\ref{colla}
\bea
\langle 0|O_c^a|0 \rangle  &=&  2 \sin^2 \theta_W \Big \{\delta_{ \tilde A 3} \delta_{\tilde B 3} + \{\delta_{\tilde A \tilde B}-3\delta_{ \tilde A 3} \delta_{\tilde B 3}\}\frac{\alpha_W}{\pi}\{2 \log(\frac{M_{\chi}}{\nu}) \log(\frac{\mu}{M_W})\} \nn\\
&+& \delta_{ \tilde A 3} \delta_{\tilde B 3}\frac{\alpha_W}{2\pi}\beta_0 \log(\frac{\mu}{M_W})+(\delta_{ \tilde A \tilde B} -\delta_{\tilde A 3}\delta_{\tilde B 3})\frac{\alpha_W}{\pi}\int_{z_{cut}}^1 dz \, P_{gg}^*(z)\log(\frac{\mu}{M_W})\Big \}
\eea
where,
\bea
 P_{gg}^*(z) &=& 2\left[ z(1-z)+ \frac{z}{(1-z)^+}+\frac{1-z}{z}\right],
\eea
with $z$ the energy fraction of $M_{\chi}$ carried by the photon and $z_{cut}$ is the threshold value for the photon to be detected by the experiment.  For HESS, $z_{cut}$ varies between 0.89 to 0.83 over the range of masses 0.5 -19 TeV., 
so we adopt an average value of $z_{cut} =$ 0.85.  The dependence on $z_{cut}$ appears because of the semi-inclusive nature of our calculation.  We are implicitly including three-body processes where the observed photon has lost a finite amount of its energy.  In a forthcoming work \cite{futuremv}, we will calculate the dependence of the rate on the possibly numerically important quantity $\log (1-z_{cut})$.  Finally, $\beta_0 =\frac{19}{6}$ is the one-loop SU(2) $\beta$-function coefficient. 

The operator $O_s^b$ has a trivial structure and hence does not receive radiative corrections, meaning its anomalous dimension is 0.
At one loop, the diagrams that contribute to the $O_c^b$ matrix element are the same as those for $O_c^a$ in Fig.~\ref{colla}. However, $O_c^b$ is a color singlet, meaning the real and virtual double poles cancel, so it does not have any double logs.  Nonetheless, it does have a non-cusp, one-loop anomalous dimension.
\bea
\langle 0|O_s^b|0 \rangle  &=& \delta_{\tilde A \tilde B}\delta_{A' B'}\nn\\
\langle 0|O_c^b|0 \rangle  &=&  2 \sin^2 \theta_W \delta_{ \tilde A \tilde B}\Big \{ 1+ \frac{\alpha_W}{\pi}\left(\frac{\beta_0}{2}+2\int_{z_{cut}}^1 dz \, P_{gg}^*(z)\right)\log(\frac{\mu}{M_W})\Big \}
\eea

Since our operators mix, we work with a matrix of anomalous dimensions,
\beq \mu \frac{d}{d\mu}\left(\begin{array}{c} O^{c,s}_a \\ O^{c,s}_b \end{array}\right)=\left( \begin{array}{cc} \gamma^{c,s}_{\mu,aa} & \gamma^{c,s}_{\mu,ab}  \\ \gamma^{c,s}_{\mu,ba} & \gamma^{c,s}_{\mu,bb} \end{array} \right) \left(\begin{array}{c} O^{c,s}_a \\ O^{c,s}_b \end{array}\right). \eeq
\beq \nu \frac{d}{d\nu}\left(\begin{array}{c} O^{c,s}_a \\ O^{c,s}_b \end{array}\right)=\left( \begin{array}{cc} \gamma^{c,s}_{\nu,aa} & \gamma^{c,s}_{\nu,ab}  \\  \gamma^{c,s}_{\nu,ba} & \gamma^{c,s}_{\nu,bb}\end{array} \right) \left(\begin{array}{c} O^{c,s}_a \\ O^{c,s}_b \end{array}\right), \eeq
given by
\begin{align}
\gamma^c_{\mu,aa} &= \frac{3\alpha_W}{\pi} \log(\frac{\nu^2}{4M^2_\chi})+\frac{\alpha_W}{2\pi}\left(\beta_0-2\int_{z_{cut}}^1 dz \, P_{gg}^*(z) \right), \nn \\
\gamma^c_{\mu,ab} &= -\frac{\alpha_W}{\pi} \log(\frac{\nu^2}{4M^2_\chi})+\frac{\alpha_W}{\pi}\int_{z_{cut}}^1 dz \, P_{gg}^*(z), \nn \\
\gamma^s_{\mu,aa} &= -\frac{3\alpha_W}{\pi} \log(\frac{\nu^2}{\mu^2})+\frac{3\alpha_W}{\pi}, \;\;\; \gamma^s_{\mu,ab} = \frac{\alpha_W}{\pi} \log(\frac{\nu^2}{\mu^2})-\frac{\alpha_W}{\pi}. \nn \\
\gamma^c_{\mu,bb} &= \frac{\alpha_W}{\pi}\left(\frac{\beta_0}{2}+2\int_{z_{cut}}^1 dz \, P_{gg}^*(z)\right).
\label{ads}
\end{align}
Although they will play no further role in our calculation as our soft and collinear logs are minimized by running $\mu$ to $M_W$, for completeness we include our $\nu$-anomalous dimensions,
\bea
\gamma^c_{\nu,aa} &=& \frac{3\alpha_W}{\pi} \log(\frac{\mu^2}{M_W^2}), \;\;\; \gamma^s_{\nu,aa} = -\frac{3\alpha_W}{\pi} \log(\frac{\mu^2}{M_W^2}), \nn \\
\gamma^c_{\nu,ab} &=& -\frac{\alpha_W}{\pi} \log(\frac{\mu^2}{M_W^2}), \;\; \gamma^s_{\nu,ab} = \frac{\alpha_W}{\pi} \log(\frac{\mu^2}{M_W^2}). 
\label{ads2}
\eea
The terms in the RG matrix which are not explicitly stated are all 0.

\subsection{Resummed Wino Annihilation Rate}

We use the RG invariance of the cross section to obtain equations for the Wilson coefficients of the operators in Eq.~\ref{eq:winoops},
\bea
\mu \frac{d}{d\mu}C_{2,4}(\mu) &=& - (\gamma^c_{\mu,aa}+\gamma^s_{\mu,aa}) \, C_{2,4} \nn \\
\mu \frac{d}{d\mu}C_{1,3}(\mu) &=& - (\gamma^c_{\mu,ba}+\gamma^s_{\mu,ba}) \, C_{2,4} - \gamma_{\mu,bb}^c \, C_{1,3}. 
\label{wilsonrg}
\eea
As discussed above, since the running of the Wilson coefficients combines both soft and collinear sectors, it must be independent of the rapidity scale $\nu$.  Plugging in the results of Eq.~\ref{ads}, we see that this is the case.
The soft and collinear  sectors have no large logs if we choose the $(\mu,\nu)$ scales to be $(M_W,M_W)$ and
 $(M_\chi,M_W)$, respectively. At leading double log accuracy we can resum all of the relevant terms
 by choosing $\mu=M_W$. In this case all the large logs reside in the renormalized parameter $C_{i}(\mu=M_W)$
 and the rapidity running may be neglected.
 
Our objective is to perform a controlled calculation of the WIMP annihilation rate, including the important nonperturbative effects of Sommerfeld enhancement and Sudakov resummation.  In our earlier papers, we focused on leading log (LL) resummation since at the thermal relic mass of 3 TeV, $\alpha_W \log^2(\frac{2M_{\chi}}{M_W}) \approx$ 0.6, and over the range of the HESS experiment (up to around 18 TeV), this quantity grows larger than 1 \cite{Baumgart:2014vma,Baumgart:2014saa}. Over the range probed by HESS, single-log corrections, $\alpha_W\log(\frac{2M_{\chi}}{M_W}) \sim$ 10-20\%, which is non-negligible.  Thus, we include them at fixed order, giving our LL$^\prime$ result.  Therefore, we are accounting for all the terms of the form $\alpha^{n+1}_W\log^{2n+1}(\frac{2M_{\chi}}{M_W}) $. To do a full NLL resummation, one would need to include the two-loop cusp anomalous dimension as well.  Given the percent-scale effect this has, this is beyond the order to which we are working.  

We can easily write down the solutions using the boundary conditions $C_2(2M_{\chi}) = -2C_1(2M_{\chi}),  C_4(2M_{\chi}) = C_1(2M_{\chi}), C_3(2M_{\chi}) = 0$ \cite{Baumgart:2014vma,Baumgart:2014saa},
\bea
C_1(\mu) &= &\frac{1}{3}C_1(2M_{\chi}) \left[ \left(1 + P_g L \right) + 2 E_1 \left(1 + P'_g L - \frac{\alpha_W^2}{\pi^2} \beta_0 L^3\right) \right] \nn \\
C_2(\mu) &=& -2C_1(2M_{\chi}) E_1 \left(1 + P_g' L - \frac{\alpha_W^2}{\pi^2}\beta_0L^3\right)\nn\\
C_3(\mu) &= & \frac{1}{3} C_1(2M_{\chi}) \left[ \left(1 + P_g L\right) - E_1 \left(1 + P'_g L - \frac{\alpha_W^2}{\pi^2} \beta_0 L^3\right) \right] \nn\\
C_4(\mu) &=& C_1(2 M_{\chi}) E_1 \left(1 + P'_g L - \frac{\alpha_W^2}{\pi^2} \beta_0 L^3\right)
\label{eq:winowc}
\eea
where 
\bea
P_g &=& \frac{\alpha_W}{\pi} \left(\frac{\beta_0}{2} +2 \int_{z_{cut}}^1 dz\, P_{gg}^*(z)\right), \;\;\; P'_g = \frac{\alpha_W}{\pi}\left(\frac{\beta_0}{2} - \int_{z_{cut}}^1 dz\, P_{gg}^*(z)+3\right) \nn \\
E_1&=& \exp \left[ -\frac{3\alpha_W}{\pi} \log^2 \left( \frac{2M_{\chi}}{M_W} \right) \right], \ \ \ L= \log \left( \frac{2M_{\chi}}{M_W} \right).
\label{eq:winoauxdef}
\eea
To write down the final cross section, we need to evaluate the photon fragmentation function, Eq.~\ref{eq:fragdefw}, at the scale $M_W$.  At LL$^\prime$, the photon can fragment into charged SM states,  so $F^{\gamma}_{\tilde A\tilde B}(M_W)$ is no longer its tree level value.  In particular, this correction enters via the gauge-boson wavefunction renormalization, proportional to the corresponding $\beta$-function coefficient.  Since this includes light fermionic states with mass lower than $M_W$, we therefore modify our result to take into account these corrections in the following way 
\bea
\frac{\alpha_W}{2\pi} \beta_0 \log(\frac{2M_{\chi}}{M_W}) \rightarrow  \frac{\alpha_W}{2\pi} \beta_0 \log(\frac{2M_{\chi}}{M_W}) +\Pi_{\gamma \gamma},
\label{eq:pwrconv}
\eea
and similarly for hypercharge, where $\Pi_{\gamma \gamma}$ ($\equiv \frac{\alpha(0)}{\alpha(M_Z)}$-1) is the photon self energy function evaluated at the scale $M_Z$ and $\alpha$ is the fine-structure constant. The value of this quantity is derived from  experiment to be -0.0594  \cite{Burkhardt:2001xp}. We refer the reader to  Appendix \ref{sec:pwr} for details.

Following the discussion in \cite{Baumgart:2014saa}, which fixes the tree-level matching coefficient as
\beq
C_1(M_{\chi}) = \frac{\pi \alpha_W^2 \sin^2\theta_W}{2M_{\chi}^3},
\label{eq:tlmatch}
\eeq
we write the final cross-section in terms of the Sommerfeld enhancement factors,
\bea
\sigma  v &=&  \frac{\pi \alpha_W^2 \sin^2\theta_W}{8M_{\chi}^4}  
\left\{ \frac{4}{3} f_-^\prime |\psi_{00}(0)|^2 
+ 4 f_+^\prime \mid \!\psi_{\pm}(0)\!\mid^2 \right. \nn \\
&&+ \left. \frac{4}{3} f_-^\prime \left(\psi_{00}(0)\psi^*_{\pm}(0)+{\rm h.c.}\right) 
\right \},
\label{eq:diffrate}
\eea
where
\bea
f_\pm^\prime &=& (1 \pm E_1) +\left( L P_g+\Pi_{\gamma \gamma}\right) \pm E_1\left( P_g'L+\Pi_{\gamma \gamma}  \right)\nn\\
&&\mp  2E_1\frac{\alpha_W}{\pi} L^2\left(\frac{\alpha_W}{2\pi} \beta_0 L +\Pi_{\gamma \gamma}\right).
\label{eq:sud}
\eea
We define the nonrelativistic wavefunctions as
\begin{eqnarray}
\psi_{00}(0) &=& \langle 0 | (\chi^0)^\top  i \sigma^2 \chi^0 |\chi^0 \chi^0 \rangle_S = 2\sqrt{2} M_\chi \, \psi_1(0) \nn \\
\psi_{\pm}(0) &=& \langle 0 |  (\chi^-)^\top i \sigma^2 \chi^+ | \chi^0 \chi^0 \rangle_S = 2 M_\chi \, \psi_2(0),
\label{eq:wfxn}
\end{eqnarray}
where $|\chi^0 \chi^0 \rangle_S = \frac{1}{\sqrt{2}} (|\chi^0_{\uparrow}(p_1) \chi^0_{\downarrow}(p_2)\rangle-|\chi^0_{\downarrow}(p_1) \chi^0_{\uparrow}(p_2)\rangle)$, a spin singlet state.  The factors $\psi_{1,2}$ in Eq.~\ref{eq:wfxn} are the dimensionless Sommerfeld enhancement factors.  We briefly sketch their calculation in Appendix \ref{sec:se}, with more details in \cite{Hisano:2004ds,Cohen:2013ama,Baumgart:2014saa}.  We note that the LL$^\prime$ rate is schematically the same as LL, the only difference coming from the correction to the Sudakov factors in Eq.~\ref{eq:sud}.  We recover the earlier result by setting $L,\Pi_{\gamma \gamma}$ to 0 in that formula.

\section{Higgsino}
\label{sec:hino}

We next turn to the case of higgsino-like dark matter.  Just as with the wino, one can take this state to be the neutralino LSP, in the limit where mixing with gauginos is negligible.  However, it could arise in a simplified model of Dark Matter, where we augment the SM by two SU(2) doublets with hypercharge $\pm \frac{1}{2}$.  The one complication in this sector is that direct detection constraints prevent the DM to from being a pure doublet \cite{Nagata:2014wma}.  We need mixing with another state (such as the gauginos) to split the Majorana components of the neutral fermion by at least 200 keV.  However, the requirements on the state being mixed are only that it have mass less than 10$^9$ GeV, which is sufficiently decoupled that in practice we work with a doublet possessing splittings between its neutral components and between the neutral and charged fermions.  These mass differences are much more model-dependent than that of the wino.  We thus study both a narrowly-split limit, just allowed by direct detection constraints, and a more widely split scenario which could arise in the MSSM neutralino sector from having gaugino soft masses $M_{1,2}$ a factor of a few greater than $|\mu|$.

Prior to dressing with soft Wilson-lines, the minimal operator basis that we need is
\bea
O_1 &=& g^2 g'^2 \left[ (\bar \chi \, \gamma^5 \tau^a \chi) (\bar \chi \, \gamma^5 \tau^a \chi) + \frac{\tan \theta^2_W}{4} \, (\bar \chi \, \gamma^5 \chi) (\bar \chi \, \gamma^5 \chi) \right] \, B \, B \nn \\
O_2 &=& \frac{g^4}{4} (\bar \chi \, \gamma^5 \chi) (\bar \chi \, \gamma^5 \chi) \, B^A \, B^A \nn \\
O_3 &=& g^2 g'^2 \, (\bar \chi \, \gamma^5 \tau^A \chi) (\bar \chi \, \gamma^5 \tau^B \chi) \, B^A \, B^B \nn \\
O_4 &=& \left( \frac{g^3 g' + g g'^3}{2} \right) \left[ (\bar \chi \, \gamma^5 \tau^A \chi) (\bar \chi \, \gamma^5 \chi) \, B^A \, B + (\bar \chi \, \gamma^5 \chi) (\bar \chi \, \gamma^5 \tau^A \chi) \, B \, B^A \right] \nn \\
O_5 &=& (\bar \chi \, \gamma^5 \tau^A \chi) (\bar \chi \, \gamma^5 \tau^A \chi) \, B^B \, B^B.
\label{eq:hinoops}
\eea
$B^A$ is the SU(2) gauge-boson field (implicitly the $\perp$ portion of $B^A_n$) while $B_n$ is the hypercharge field.  The coupling-dependent prefactors arise from matching the 2$\rightarrow$2 WIMP annihilation to gauge bosons at tree-level.  This gives dimension-five, $(\bar \chi \chi) BB$ operators, with which we then perform an OPE ({\it cf.}~Fig.~\ref{semi}) to obtain operators for the semi-inclusive annihilation rate to $\gamma + X$.  Normalizing the operators in this way automatically gives 0 when we add up $O_{1 \textendash 4}$ ($O_5$ is not generated at tree level), as it should, when we set the fermions to their neutral components and project the gauge bosons onto photons, $B^A \rightarrow \sin \theta_W \, \gamma$, $B \rightarrow \cos \theta_W \, \gamma$.  Thus, just as in the wino sector, we will just need a nonzero tree-level process, like $\chi^+ \chi^- \rightarrow \gamma \gamma + \frac{1}{2} \gamma Z$, to fix a common Wilson coefficient.\footnote{Because of the identity $\tau^a_{ij} \tau^a_{i'j'} = \frac{1}{2}\delta_{ij'}\delta_{i'j} - \frac{1}{4} \delta_{ij} \delta_{i'j'}$, the basis in Eq.~\ref{eq:hinoops} is not orthogonal.  For simplicity though, we have chosen to contract color indices between fermions on the same side of the cut.}
As explained below Eq.~\ref{eq:winoops}, we use the vacuum insertion approximation in the WIMP sector, but we have dropped the explicit vacuum projector.  Just as before, there is also an implicit Lorentz contraction and projection onto a single-photon state between the $B,B^A$ fields ({\it cf.}~Eq.~\ref{eq:photproj}).  As in the wino case, since we are annihilating Majorana fermions, the only relevant bilinear is $\bar \chi \, \gamma_5 \chi$. The spin-one operators are irrelevant since Fermi statistics would lead to an antisymmetric SU(2) initial state, and we are interested in the annihilation of two neutral particles.  Furthermore, P-wave annihilation is velocity suppressed.  In the fermion sector we have combined the two doublets into $\chi$ with,
\beq
\chi \equiv \left( \begin{array}{c}
\tilde h_u \\
\epsilon \, \tilde h^*_d
\end{array} \right), \;\;\;\;\; \bar \chi = \left( -\epsilon \, \tilde h_d \;\;  \tilde h_u^* \right),
\eeq
where $\tilde h_{u,d}$ are the higgsino gauge eigenstates.  Our operators therefore also have an intrinsic identity in doublet-space.  In the limit where the mass eigenstates are pure higgsino, we have
\bea
\chi_1^0 &=& \frac{1}{\sqrt 2} \left(h^0_d - h^0_u \right) \\
\chi_2^0 &=& -\frac{i}{\sqrt 2} \left(h^0_d + h^0_u \right),
\eea
where $\chi_1^0$ is the lighter mass state.  We also have a charged Dirac fermion, $\chi^{+\, \top} = (\tilde h_u^+ \,\, \tilde h_d^{-*})$.

We can now decompose the operators in Eq.~\ref{eq:hinoops} into the soft and collinear sectors, as in Eq.~\ref{eq:wscops}, obtaining a factorization theorem for the annihilation rate.  This is identical to our approach with winos and some of the resulting operators will even be identical, therefore receiving the same radiative corrections.  In the soft sector we get,
\bea
O^a_s &=& S_{v A^\prime A}^\top \, S_{vB B^\prime} \, S_{n  \tilde A A}^\top \, S_{n B \tilde B} \nn \\
O^b_s &=& \delta_{\tilde A \tilde B} \delta_{A' B'} \nn \\
O^c_s &=& S_{v A^\prime A}^\top \, S_{n \, \tilde A A}^\top.
\label{eq:hsops}
\eea
In terms of the basis in Eq.~\ref{eq:hinoops}, these operators dress $O_3,\, O_5,$ and $O_4$, respectively.  We see that $O^a_s$ and $O^b_s$ are identical to operators found in the wino sector and $O^c_s$ is effectively the ``square root'' of $O^a_s$.  The full operators $O_1$ and $O_2$ contain various singlet structures, some of which are trivial in triplet-space, but they do not receive radiative corrections and are not generated by mixing, so do we not detail their form.  Despite the higgsinos being doublets, we can obtain identical soft Wilson-line structures to the wino case by using the identity $s^\dag \tau^a \, s = \tau^c (S^\top)^{ca}$, where $s$ is the soft Wilson-line in the fundamental representation and $S$ is the adjoint.  In the collinear sector, we have
\bea
O_c^a &=& B_{\tilde A} \, B_{\tilde B} \nn \\
O_c^b &=& B_D \, B_D \, \delta_{\tilde A \tilde B}. \nn \\
O_c^c &=& B_{\tilde A} \, B + B B_{\tilde A} \nn \\
O_c^d &=& BB,
\label{eq:hcops}
\eea
where here $O_c^{a,b}$ are repeats from the wino sector.\footnote{Strictly speaking, $O_2$ in Eq.~\ref{eq:hinoops} contains a fifth collinear structure, which we could call $O_c^e = B_D \, B_D$.  It has no additional $\delta$, unlike $O_c^b$, since the rest of the operator has trivial triplet structure.  The loop corrections to these operators are identical, up to the overall $\delta$.  We can therefore recycle our $O_c^b$ calculation for $O_c^e$, and neither generates other structures.  The running of $O_c^a$ though, will generate contributions to $O_c^b$, but not $O_c^e$.}

The details of our factorization theorem, which is identical to the wino result, can be found in \cite{Baumgart:2014vma,Baumgart:2014saa}.  Because of the more involved operator structure, we will also decompose the operators in Eq.~\ref{eq:hinoops} into the WIMP sector, defining $O_i^{\rm NR}$ to be the $\chi$-field component of the full operator $O_i$, including any Dirac and color matrices that contract with the fermions.  For example,
\beq
O_{3 \, \rm NR}^{AB} = (\bar \chi \, \gamma^5 \tau^A \chi) (\bar \chi \, \gamma^5 \tau^B \chi).
\eeq
The annihilation spectrum may be written as
\bea
\label{cross}
&& \frac{1}{E_\gamma}  \frac{d\sigma}{dE_\gamma} = \frac{1}{4M_\chi^2 v}  \Bigg \{\langle 0 | O^a_s | 0 \rangle \Bigg[ \int d n \cdot p \, \Bigg\{ C_3(M_\chi, n \cdot p) \langle p_1 p_2 \mid O_{3 \, \rm NR}^{A'B'} \mid p_1 p_2 \rangle \Bigg\} F^\gamma_{\tilde A \tilde B}\left( \frac{2E_\gamma}{n \cdot p} \right) \Bigg] \nn\\
&&+  \Bigg[ \int d n \cdot p \,\Bigg\{  C_2(M_\chi, n \cdot p)\langle p_1 p_2 \mid O_{2 \, \rm NR} \mid p_1 p_2 \rangle 
+ C_5(M_\chi, n \cdot p)\langle p_1 p_2 \mid O_{5 \, \rm NR} \mid p_1 p_2 \rangle \Bigg\} F^\gamma_{\tilde A\tilde A}\left( \frac{2E_\gamma}{n \cdot p} \right)  \Bigg]\nn\\
&+& \langle 0 | O^c_s | 0 \rangle \Bigg[ \int d n \cdot p \,\Bigg\{ C_4(M_\chi, n \cdot p)\langle p_1 p_2 \mid O_{4 \, \rm NR}^{A'} \mid p_1 p_2 \rangle \Bigg\} F^\gamma_{\tilde A}\left( \frac{2E_\gamma}{n \cdot p} \right)  \Bigg]\nn\\
&+& \Bigg[ \int d n \cdot p \,\Bigg\{ C_1(M_\chi, n \cdot p)\langle p_1 p_2 \mid O_{1 \, \rm NR} \mid p_1 p_2 \rangle \Bigg\} F^\gamma \left( \frac{2E_\gamma}{n \cdot p} \right)  \Bigg] \Bigg \},
\label{hcross}
\eea
where  
$F^\gamma_{\tilde A \tilde B},F^\gamma_{\tilde A},F^\gamma$ are fragmentation functions defined by
\bea
F^\gamma_{\tilde A \tilde B}\left( \frac{n\cdot k}{n\cdot p} \right) &=& \int \frac{ dx_-}{2\pi}e^{in \cdot p x_-}\langle 0 \mid B^{\perp \mu}_{n,\, \tilde A}(x_-) \mid \gamma(k_n)+X_n \rangle \nn \\
&\times& \langle \gamma(k_n)+X_n \mid  B^\perp _{n,\, \mu \tilde B}(0)  \mid 0 \rangle \nn\\
F^\gamma_{\tilde A}\left( \frac{n\cdot k}{n\cdot p} \right) &=& \int \frac{ dx_-}{2\pi}e^{in \cdot p x_-}\langle 0 \mid B^{\perp \mu}_{n,\, \tilde A}(x_-) \mid \gamma(k_n)+X_n \rangle \nn \\
&\times& \langle \gamma(k_n)+X_n \mid  B^\perp _{n,\, \mu}(0)  \mid 0 \rangle \nn\\
F^\gamma\left( \frac{n\cdot k}{n\cdot p} \right) &=& \int \frac{ dx_-}{2\pi}e^{in \cdot p x_-}\langle 0 \mid B^{\perp \mu}_n (x_-) \mid \gamma(k_n)+X_n \rangle \nn \\
&\times& \langle \gamma(k_n)+X_n \mid  B^\perp_{n,\, \mu} (0)  \mid 0 \rangle ,
\label{eq:fragdef}
\eea 
$C_{1\textendash 4}$ are the matching coefficients that give the probability for the dark matter to annihilate with total +-momentum in the $n$-direction of $ n \cdot p$.
$F^\gamma$ and $F^\gamma_{\tilde A \tilde A}$ are the canonical fragmentation functions giving the probability of an initial boson with momentum $p$ to yield
a photon with momentum fraction $n\cdot k /n \cdot p$ after splitting.  Just as with the wino, we get gauge nonsinglet fragmentation functions, which are rendered physical by electroweak symmetry breaking.   Since the contributions in Eq.~\ref{hcross} proportional to $F^\gamma_{\tilde A \tilde A}$ are not sensitive to the nonsinglet nature of the initial state, they will only contribute large double logs from mixing with $O_3$.

\subsection{Anomalous Dimensions}
\label{sec:ad}

We now calculate the anomalous dimensions for the soft and collinear sector operators in Eqs.~\ref{eq:hsops}, \ref{eq:hcops}. 
Four of the operators $O_{c,s}^a,O_{c,s}^b$ are exactly the same as the wino case. However, due to the coupling of the higgsino to the Hypercharge field, we obtain three new nontrivial structures, $O^c_{c,s}$, contained in $O_3$, and $O^c_d$ in $O_4$, which we repeat here,
\bea
O^c_s &=& S_{v A^\prime A}^\top S_{n \,  \tilde A A}^\top \nn \\
O^c_c &=& B_{\tilde A} \, B + B B_{\tilde A} \nn \\
O^d_c &=& B\, B
\eea
\begin{figure}
\centerline{\scalebox{0.5}{\includegraphics{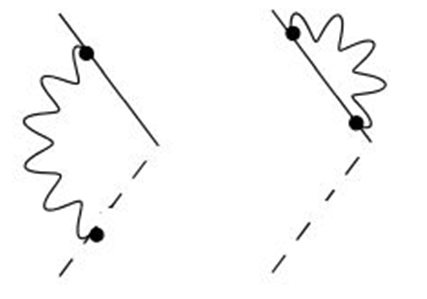}}}
\vskip-0.2cm
\caption[1]{The diagrams contributing to $O^c_s$. The solid/dashed lines indicate time/light-like Wilson lines. }
\label{softb} 
\end{figure}
\begin{figure}
\centerline{\scalebox{0.55}{\includegraphics{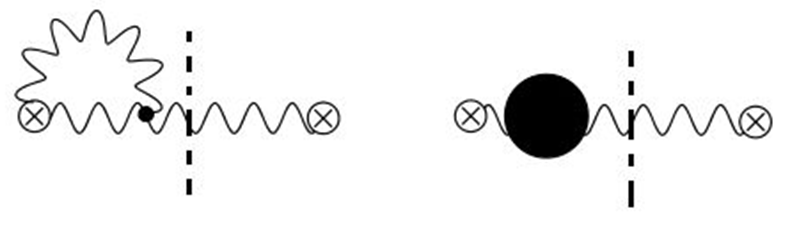}}}
\vskip-0.2cm
\caption[1]{The diagrams contributing to $O^c_c$. The dark blob contains fermion, Higgs and gauge boson loops. }
\label{collc} 
\end{figure}
The operator $O^c_s$ has a unique color structure, so that it mixes only into itself (Fig.\ref{softb}). $O^c_c$ contributes to a hybrid Weak-Hypercharge fragmentation function, which is why it gets only virtual corrections (Fig. \ref{collc}). The hypercharge field is Abelian, so it does not get any Wilson-line contributions.  
\bea
\langle 0|O^c_s|0 \rangle  &=& \delta_{A' \tilde A}\left[ 1- \frac{\alpha_W}{\pi}\left\{2\log(\frac{\nu}{\mu}
)\log(\frac{\mu}{M_W})+\log^2(\frac{\mu}{M_W})-\log(\frac{\mu}{M_W})\right\}\right]\nn\\
\langle 0|O^c_c|0 \rangle  &=& 2 \sin \theta_W \cos \theta_W \delta_{ \tilde A 3}\left[1-\frac{\alpha_W}{\pi}\{2 \log(\frac{M_{\chi}}{\nu}) \log(\frac{\mu}{M_W})\}+\frac{\alpha_W}{4\pi}\beta_0\log(\frac{\mu}{M_W})+\frac{\alpha'}{4\pi}\beta_0'\log(\frac{\mu}{M_W})\right], \nn\\
\eea
where $\beta_0'= -\frac{41}{6}$ is the leading order hypercharge beta function coefficient. 
\begin{figure}
\centerline{\scalebox{0.55}{\includegraphics{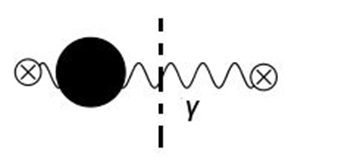}}}
\vskip-0.2cm
\caption[1]{The diagrams contributing to $O^d_c$. The dark blob contains fermion and Higgs loops. }
\label{colld} 
\end{figure}
$O^d_c$, which contributes to the hypercharge fragmentation function, only gets wavefunction renormalization corrections (Fig. \ref{colld}), which we discuss further in Appendix \ref{sec:pwr}.  
\bea
\langle 0|O^d_c|0 \rangle  &=& 2 \cos^2 \theta_W \left[1+\frac{\alpha'}{2\pi}\beta_0'\log(\frac{\mu}{M_W})\right] 
\eea

From these matrix elements, we can compute the regular and rapidity anomalous dimensions.  None of the new contributions generates any mixing, so we simply get,
\bea
\mu \frac{d}{d \mu} O^c_c &=& \gamma_{\mu\, c}^c \, O^c_c ,\;\;\; \nu \frac{d}{d \nu} O^c_c = \gamma_{\nu\, c}^c \, O^c_c \nn\\
 \mu \frac{d}{d \mu} O_s^c &=& \gamma_{\mu\, c}^s \, O^c_s, \;\;\; \nu \frac{d}{d \nu} O^c_s = \gamma_{\nu\, c}^s \, O_s \nn\\
 \mu \frac{d}{d \mu} O^d_c &=& \gamma^c_{\mu\, d}\, O^d_c \nn\\
\eea
and
\bea
 \gamma_{\mu \,c}^c &=& \frac{\alpha_W}{\pi}\left(\log(\frac{\nu^2}{4M_{\chi}^2})+\frac{\beta_0}{4}\right) +\frac{\alpha'}{4\pi}\beta_0', \;\;\; \gamma_{\nu \, c}^c= \frac{\alpha_W}{\pi} \log(\frac{\mu^2}{M_W^2})\nn\\
 \gamma_{\mu\, c}^s &=& -\frac{\alpha_W}{\pi}\log(\frac{\nu^2}{\mu^2})+\frac{\alpha_W}{\pi}, \;\;\; \gamma_{\nu\, c}^s= -\frac{\alpha_W}{\pi} \log(\frac{\mu^2}{M_W^2})\nn\\
\gamma_{\mu\, d}^c &=& \frac{\alpha'}{2\pi}\beta_0'
\eea
The terms not explicitly stated are all 0.  We note that $O^d_c$ has no rapidity anomalous dimension as it receives contributions only at the single-log level.  We trivially see that $\nu$-dependence drops out, as it must, when we add $\gamma_{\mu \,c}^c + \gamma_{\mu \,c}^s$, the combination that will appear in running the Wilson coefficient, $C_4$, for the operator in Eq.~\ref{eq:hinoops}.  Also, we see that both soft and collinear rapidity anomalous dimensions will vanish when we set $\mu = m_W$, so we will not need the rapidity RG.

\subsection{Resummed Higgsino Annihilation Rate}

We can now use RG invariance of the cross section to compute the RG equations for the Wilson coefficients 
\bea
\mu \frac{d}{d \mu} C_1 &=& - \gamma_{\mu \, d}^c \, C_1 \nn \\
\mu \frac{d}{d \mu} C_2 &=& - \gamma_{\mu,bb}^c \, C_2 \nn \\
\mu \frac{d}{d \mu} C_3 &=& - \left(\gamma_{\mu,aa}^c+\gamma_{\mu,aa}^s\right) C_3 \nn \\
\mu \frac{d}{d \mu} C_4 &=& - \left(\gamma_{\mu \, c}^c+\gamma_{\mu \, c}^s\right) C_4 \nn \\
\mu \frac{d}{d \mu} C_5 &=& -\frac{1}{3} \left(\gamma_{\mu,ab}^c+\gamma_{\mu,ab}^s\right) C_3 - \gamma_{\mu,bb}^c \, C_5
\label{RG}
\eea
As in the wino case, our objective is to obtain the LL$^\prime$ cross section in order to have a controlled calculation for higgsino annihilation. For the thermal relic mass of 1 TeV,  we see that the Sudakov logarithm is not so large, $\log(2M_\chi/M_W) \approx$ 3.2, so a pure LL calculation at this value may not capture all important contributions.  Tree-level matching tells us $C_1(2M_{\chi}) = C_{2 \textendash 4}(2M_{\chi}) \equiv C(2 M_\chi)$ and $C_5(2M_{\chi}) = 0$.
Using these boundary conditions and the RG equations, we can write down the solutions for the Wilson coefficients at this order,  
\bea
C_1(\mu) &=&  C(2 M_\chi) \left[1+ \frac{\alpha'}{2\pi}\beta_0' L \right] \nn \\
C_2(\mu) &=&  C(2 M_\chi) \left[1+ P_g \, L \right] \nn \\
C_3(\mu) &=&  C(2 M_\chi) E_1\left[ 1 + P'_g \, L - \frac{\alpha_W^2}{\pi^2} \beta_0 L^3 \right]  \nn \\
C_4(\mu) &=&  C(2 M_\chi) E_2\left[ 1 + \left( \frac{\alpha_W}{4\pi}(4 + \beta_0) + \frac{\alpha'}{4\pi} \beta'_0 \right) L - \frac{\alpha_W^2}{3\pi^2} \beta_0 L^3 \right]  \nn \\
C_5(\mu) &=& C(2 M_\chi) \left[  \frac{1}{3} \left(1 + P_g L \right) - \frac{1}{3} E_1 \left(1+ P_g' L - \frac{\alpha_W^2}{\pi^2} \beta_0 L^3 \right) \right],
\label{eq:hinowc}
\eea
where we use the same functions, as in Eq.~\ref{eq:winoauxdef}, which we repeat here, along with the new $E_2$,
\bea
P_g &=& \frac{\alpha_W}{\pi} \left(\frac{\beta_0}{2} +2 \int_{z_{cut}}^1 dz\, P_{gg}^*(z)\right), \;\;\; P'_g = \frac{\alpha_W}{\pi}\left(\frac{\beta_0}{2} - \int_{z_{cut}}^1 dz\, P_{gg}^*(z)+3\right) \nn \\
E_1&=& \exp \left[ -\frac{3\alpha_W}{\pi} \log^2 \left( \frac{2M_{\chi}}{M_W} \right) \right], \ \ \ L= \log \left( \frac{2M_{\chi}}{M_W} \right) \nn \\
E_2&=& \exp \left[ -\frac{\alpha_W}{\pi} \log^2 \left( \frac{2M_{\chi}}{M_W} \right) \right].
\label{eq:hinoauxdef}
\eea
The cross section can now be obtained by evaluating the effective theory matrix elements at their natural scale $\mu \sim M_W$. 

In order to get a numerical result, we must perform the tree-level matching to determine $C(2 M_\chi)$, which we again do by computing $\chi^+ \chi^- \rightarrow \gamma \gamma + \frac{1}{2} \gamma Z$, which at this order is $\chi^+ \chi^- \rightarrow \gamma + X$.  The leading order cross section is 
\beq
\sigma v = \frac{\pi \alpha_W \alpha'}{4M_{\chi}^2}.
\eeq
For both this matching and our later determination of the neutral WIMP annihilation rate, it is useful to write our fermion bilinears in terms of mass eigenstates.  For simplicity, we drop the $\gamma^5$ common to all terms in this calculation.
\bea
\bar \chi \chi &=& \frac{1}{2} \, \bar \chi_1^0 \, \chi_1^0 + \frac{1}{2} \, \bar \chi_2^0 \, \chi_2^0 + \chi^+ \chi^- \nn \\
\bar \chi \tau^3 \chi &=& -\frac{1}{4} \, \bar \chi_1^0 \, \chi_1^0 - \frac{1}{4} \, \bar \chi_2^0 \, \chi_2^0 + \frac{1}{2}  \,  \chi^+ \chi^-. 
\eea 
From the effective theory description, we calculate this particular cross section from the basis of operators in Eq.~\ref{eq:hinoops}, projecting the fermions onto the charged state and the gauge bosons onto photons,
\bea
\sigma v &=& \frac{1}{4 M_\chi^2} C(2 M_\chi) \; _S\langle \chi^+ \chi^- | \sum_{i=1}^4 O_i | \chi^+ \chi^- \rangle_S, 
\label{eq:match}
\eea
The first term on the RHS is the flux factor.  Our conventions, including the polarization sum, fix 
\beq
_S\langle \chi^+ \chi^- | \sum_{i=1}^4 O_i | \chi^+ \chi^- \rangle_S = 32\pi^2 \, \alpha_W \, \alpha' | \langle 0 | \chi^+ \gamma^5 \chi^- | \chi^+ \chi^- \rangle_S |^2, 
\eeq
and at tree level,$\langle 0| \chi^+ \gamma^5 \chi^- | \chi^+ \chi^- \rangle_S = 2\sqrt{2} M_{\chi}$.  Thus, we get\footnote{Unlike the wino calculation, we did not pull out an overall factor of $M_\chi$ from $C(2 M_\chi)$, which is why this matching coefficient is dimension -2, while in Eq.~\ref{eq:tlmatch} we get a quantity of dimension -3.}
\beq
C(2 M_\chi) = \frac{1}{256 \, \pi \, M_\chi^2},
\eeq
where we have absorbed the bosonic phase space, for both observed photon and the recoil particle in $C(2 M_\chi)$.

We now need to account for the Sommerfeld enhancement of the higgsino annihilation rate.  The details of the potential are given in Appendix \ref{sec:se}.  The computation is quite similar to the wino, but with an additional neutral, Majorana fermion.  We thus get, in analogy with Eq.~\ref{eq:wfxn},
\begin{eqnarray}
\psi^1_{00}(0) &=& \langle 0 | (\chi_1^0)^\top  i \sigma^2 \chi_1^0 | \chi_1^0 \chi_1^0 \rangle_S = 2\sqrt{2} M_\chi \, \psi_1(0) \nn \\
\psi^2_{00}(0) &=& \langle 0 | (\chi_2^0)^\top  i \sigma^2 \chi_2^0 | \chi_1^0 \chi_1^0 \rangle_S = 2\sqrt{2} M_\chi \, \psi_2(0) \nn \\
\psi_{\pm}(0) &=& \langle 0 |  (\chi^-)^\top i \sigma^2 \chi^+ | \chi_1^0 \chi_1^0 \rangle_S = 2 M_\chi \, \psi_3(0),
\label{eq:hwfxn}
\end{eqnarray}
where in addition to having an overlap between the asymptotic, $| \chi_1^0 \chi_1^0 \rangle_S$ state and annihilation of the chargino at the origin, captured by $\psi_\pm$, we also account for the possibility that the short-distance annihilation occurs via the excited neutral state $\chi_2^0$.  The functions $\psi_{1 \textendash 3}(0)$ are the dimensionless Sommerfeld factors obtained from solving the potential in Appendix \ref{sec:se}.\footnote{For consistency with the wino notation, we have written the wavefunctions in terms of two-component fermions.  It is straightforward to convert back and forth between these and the four-component formalism we have used elsewhere.}
We can now include the results of running the Wilson coefficients (Eq.~\ref{eq:hinowc}) to get the annihilation rate for $\chi_1^0 \chi_1^0 \rightarrow \gamma + X$,
\bea
 \sigma v&=& \frac{\pi \, \alpha_W\alpha'}{16 \, M_{\chi}^4} \Bigg [ \frac{1}{4} \left[ |\psi^1_{00}(0)|^2 + |\psi^2_{00}(0)|^2 +\left(\psi^1_{00} \, \psi^{2*}_{00} + {\rm c.c.} \right)  \right] \times \nn \\
&& \Big\{ (1-E_2) \left \{ 1 + \left(\frac{\alpha_W}{\pi}+\frac{\alpha_W }{4\pi}\beta_0+\frac{\alpha'}{4\pi}\beta_0' \right)L + \Pi_{\gamma \gamma} \right\} -  \frac{s^2_W}{3}(1-E_1) (1 + P_g' L+\Pi_{\gamma \gamma}) \nn \\
&-& c_W^2\frac{\alpha_W}{\pi}\Big \{1- \int_{z_{cut}}^1 dz \, P_{gg}^*(z)\Big\}L  + (E_2 - s_W^2 E_1) \left(\frac{2\alpha_W}{3\pi}\right)L^2\left(\frac{\alpha_W \beta_0}{2\pi}L+\Pi_{\gamma \gamma}\right) \Big\} \nn \\
&+& |\psi_{\pm}(0)|^2 \times \nn\\
&& \Big\{ (1+E_2) \left \{ 1 + \left(\frac{\alpha_W}{\pi}+\frac{\alpha_W}{4\pi} \beta_0+\frac{\alpha'}{4\pi}\beta_0' \right)L + \Pi_{\gamma \gamma} \right\} -  \frac{s^2_W}{3}(1-E_1)  (1 + P_g' L+\Pi_{\gamma \gamma}) \nn \\
&-& c_W^2\frac{\alpha_W}{\pi}\Big \{1- \int_{z_{cut}}^1 dz \, P_{gg}^*(z)\Big\}L - (E_2 + s_W^2 E_1) \left(\frac{2\alpha_W}{3\pi}\right)L^2\left(\frac{\alpha_W \beta_0}{2\pi}L+\Pi_{\gamma \gamma}\right) \Big\} \nn \\
 &+& \frac{1}{2} \left( \psi^1_{00} \, \psi^{*}_{\pm} +\psi^2_{00} \, \psi^{*}_{\pm} + {\rm c.c.} \right) \times \nn \\
 &&\Big \{ \frac{s^2_W}{3}(1-E_1)  (1 + P_g' L+\Pi_{\gamma \gamma}) - s_W^2 \frac{\alpha_W}{\pi}L - (s_W^2-c_W^2) \Big \{ \frac{\alpha_W}{4\pi}\beta_0 - \frac{\alpha'}{4\pi}\beta_0' \Big\} L \nn \\
&+& c_W^2\frac{\alpha_W}{\pi} \left \{ \int_{z_{cut}}^1 dz \, P_{gg}^*(z) \right \} L + s_W^2 E_1\left(\frac{2\alpha_W}{3\pi}\right)L^2\left(\frac{\alpha_W \beta_0}{2\pi}L+\Pi_{\gamma \gamma}\right) \Big\} \Bigg ],
 \label{crossfin}
\eea
where $L= \log \left( \frac{2M_{\chi}}{M_W} \right)$ and $P_g,\,E_i,$ etc.~are given in Eq.~\ref{eq:hinoauxdef}.  

As in the case of the wino, we must account for corrections to the photon fragmentation functions Eq.~\ref{eq:fragdef} beyond tree level, including the possibility to split below the scale $M_W$ . This is seen in the above equation (\ref{crossfin}) in the $\Pi_{\gamma \gamma}$ term, which, as defined earlier is the photon self energy function at the scale $M_W$. For details, see the discussion around Eq.~\ref{eq:pwrconv} and Appendix \ref{sec:pwr}.

\section{Dark Matter Constraints and Conclusion}
\label{sec:conc}

Having calculated tree level matching, LL$^\prime$ resummation, and computed the Sommerfeld enhancement numerically, we can now evaluate the differential cross section for $\chi^0 \chi^0 \rightarrow \gamma + X$, given for winos (Eq.~\ref{eq:diffrate}) and higgsinos (Eq.~\ref{crossfin}). 

In our previous calculation at LL, we found the effect of higher order corrections to be very modest \cite{Baumgart:2014saa}.  This is in contrast to those groups that performed an exclusive two-body calculation \cite{Bauer:2014ula,Ovanesyan:2014fwa}.  We therefore wanted to examine whether the single-log contributions might be a large factor, which prompted this calculation at LL$^\prime$.  As mentioned in the Introduction, working at single-log brings in new effects which have the potential to be large: 1) photon fragmentation to pairs of SM particles, 2) real emission processes with $z<z_{cut}$.  As we now show, fragmentation is a minor effect, but accounting for finite energy fraction, $z$, is numerically important enough to warrant a follow-up study \cite{futuremv}.  As before, we adapt the exclusion analysis of \cite{Ovanesyan:2014fwa}.
  
\begin{figure}[ht!]
\centerline{\scalebox{0.2}{\includegraphics{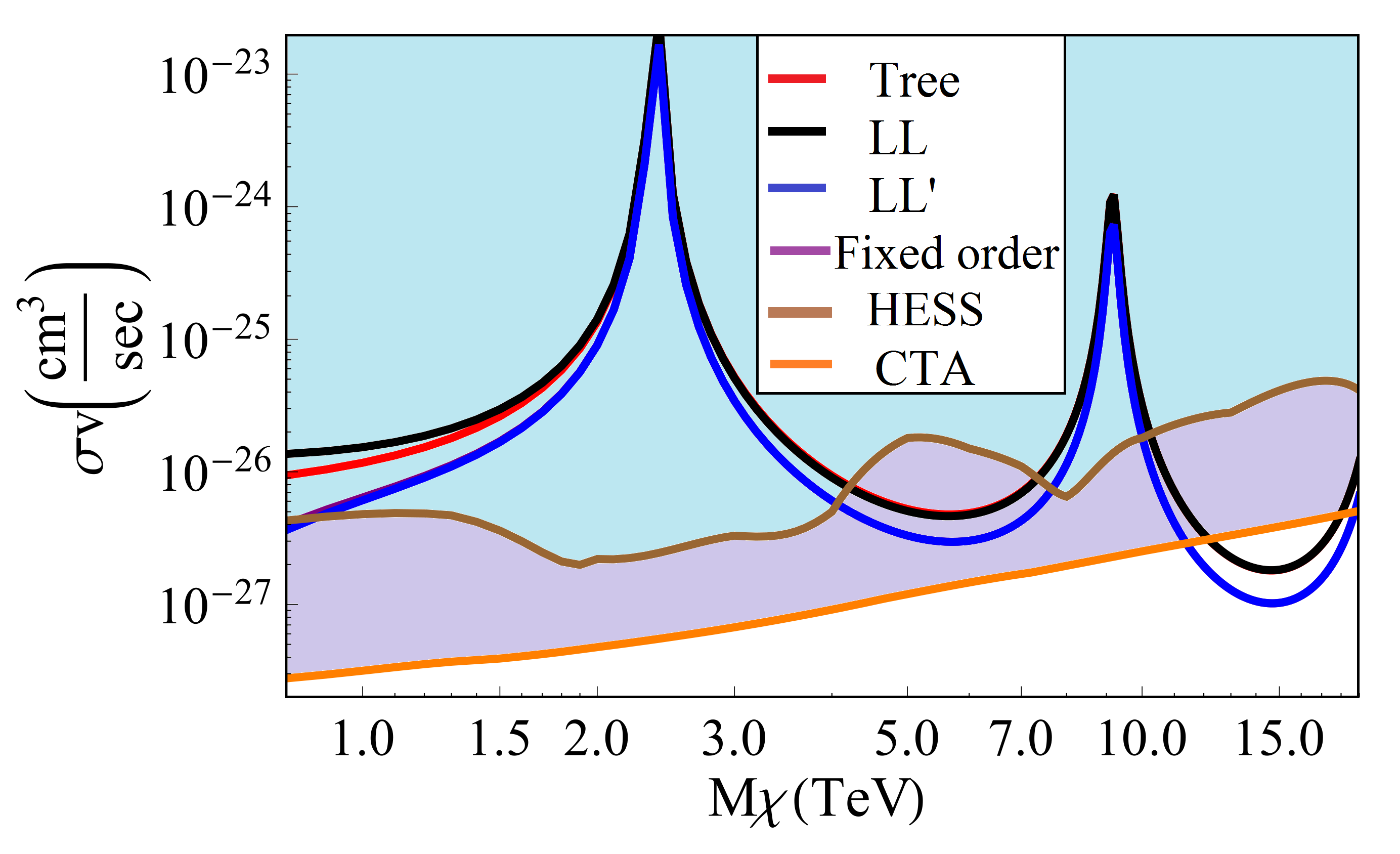}}}
\vskip-0.2cm
\caption[1]{Wino annihilation cross section at LL$^\prime$ with $z_{cut}$=0.85.}
\label{winop} 
\end{figure}

Fig.~\ref{winop} plots the wino cross section at LL$^\prime$.\footnote{The ``fixed order'' line expands the exponentials to first order and drops any other 
terms higher in $\alpha_W$.  It thus provides a quantification of the effect of summing the double logs.} This plot assumes a detector signal bin size of 0.15 $M_{\chi}$ ($z_{cut}=0.85$). We see that the single-log terms produce a noticeable correction to the LL cross section. For example, at the thermal relic mass of 3 TeV, we have 
\bea
\sigma_{\rm tree} = 5.27 \times 10^{-26} \, {\rm cm^3/s} \nn\\
\sigma_{\rm LL} = 5.07 \times 10^{-26} \, {\rm cm^3/s} \nn\\
\sigma_{\rm LL^\prime} = 3.42 \times 10^{-26} \, {\rm cm^3/s} , 
\eea
which indicates a $\sim$30 $\%$ reduction. 

\begin{figure}[ht!]
\centerline{\scalebox{0.14}{\includegraphics{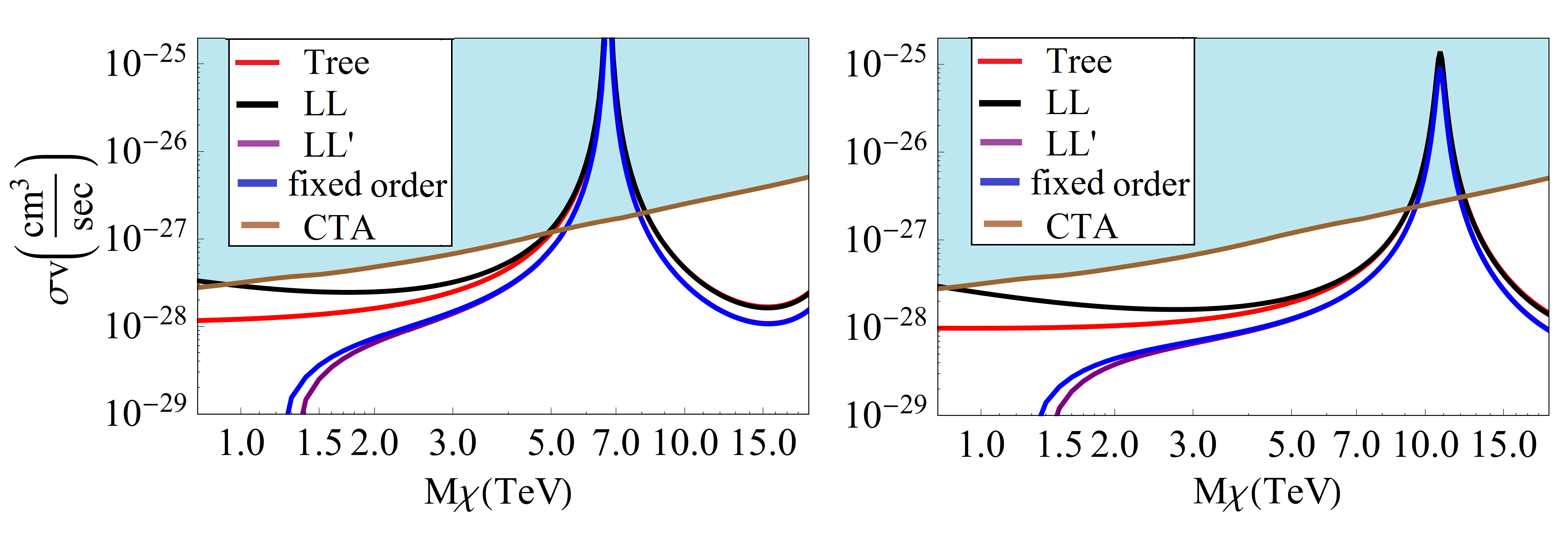}}}
\vskip-0.2cm
\caption[1]{Higgsino annihilation cross section at LL$^\prime$ with L: $\delta M_N$= 200 keV, $\delta M_+$=350 MeV, R: $\delta M_N$= 2 GeV, $\delta M_+$=480 MeV, and $z_{cut}$=0.85.}
\label{higgsino} 
\end{figure}

A similar behavior is observed in the case of the higgsinos. Here, we give plots at two limiting cases in parameter space.  The left panel of Fig.~\ref{higgsino} shows the purest doublet that direct detection constraints allow.  The neutral splitting, $\delta M_n$= 200 keV, with a chargino mass splitting $\delta M_+$= 350 MeV.  For the MSSM, this would correspond to having gaugino masses $\sim 10^8$ GeV. The right panel uses $\delta M_N$= 2 GeV and  $\delta M_+$= 480 MeV, a spectrum we would get from gaugino mass parameters just a factor of a few larger.  For $M_\chi \lesssim $ 3 TeV, we find again that the LL$^\prime$ cross section is substantially reduced. 

However, one thing to be noticed is that the full theory cross section contains single and double $\log(1-z_{cut})$, which become large in the endpoint region.  For our value of interest, $z_{cut}$=0.85, to get a handle on the importance of the endpoint logarithms, we see that $\log(1-z_{cut})^2 = \log(2 M_\chi/M_W)$ for $M_\chi \approx 1.4$ TeV.   
In fact, we see in Fig.~\ref{higgsino} that around this mass is where the LL$^\prime$ curve stops tracking LL and turns to decrease sharply, going to negative values by 1 TeV.  
The corrections arising from these endpoint logarithms (which have only been partially captured in the present form of the EFT) can no longer be ignored.  A full-theory tree-level calculation reveals that the missing terms are positive and enhance the cross section \cite{Bergstrom:2005ss}, as they must in order to restore positivity. 

\begin{figure}
\centerline{\scalebox{0.2}{\includegraphics{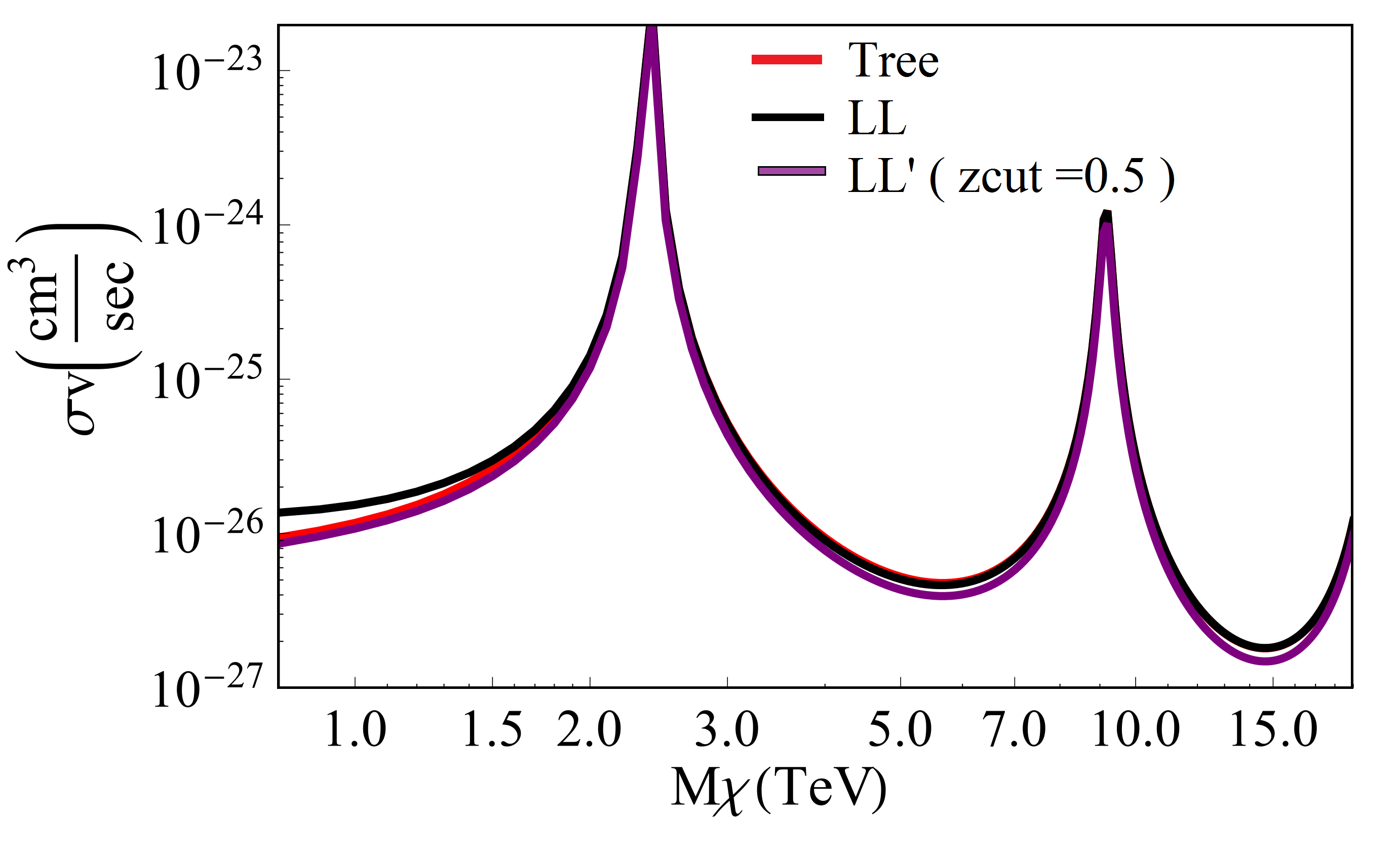}}}
\vskip-0.2cm
\caption[1]{Wino annihilation cross section at LL$^\prime$ with a detector bin size of 0.5 $M_{\chi}$  }
\label{winohalf} 
\end{figure}

To test the hypothesis that the large corrections at LL$^\prime$ arise from taking $z_{cut}$ close to the endpoint, we can take an unrealistically large ``signal bin'' with $z_{cut}$=0.5.  In this calculation, endpoint logarithms are small and we are justified in dropping them completely.  Additionally, with the larger bin, even though we smear out the photon energy at single-log order, a much greater fraction of them remain in our signal region.  We see in Fig.~\ref{winohalf} that LL$^\prime$ is indeed a small correction throughout our range.  We get a very similar result with $z_{cut}$=0.5 for higgsinos, as well.
This prompts us to conclude that as far as $\log(2M_{\chi}/M_W)$ terms are concerned, we are justified in ignoring higher order corrections, such as the two-loop cusp anomalous dimension.  However, including effects that become large in the limit of small $1-z_{cut}$ will be important going forward.
  
This exercise emphasized that it is vital to include endpoint corrections. Furthermore, as the resolution of future experiments improves, $z_{cut}$ will increase, and corrections will only get larger.  Although our results here establish the need for further study, there are two conclusions we can draw already that are worth emphasizing: 
\begin{enumerate}
\item {\bf Single logs won't save the wino:} Around the thermal relic mass of 3 TeV, $\log(1-z_{cut})^2 \lesssim \log(2 M_\chi/M_W)$, so endpoint corrections be important.  However, as we have mentioned they will be {\it positive}.  Thus, our LL$^\prime$ result represents a floor for the wino annihilation rate, and we see that this is still an order of magnitude larger than the HESS exclusion limit.  As we discussed \cite{Baumgart:2014saa}, these limits are subject to large astrophysical uncertainties, and a sufficiently large, $>1$ kpc, core can be invoked to reconcile the wino with experiment.  However, we see that the situation is qualitatively unchanged from our earlier result, and saving the wino requires a profile in tension with current results from simulation.
\item {\bf The 1 TeV thermal relic mass for the higgsino is at a tantalizing point:}  Looking at the different curves at 1 TeV in Fig.~\ref{higgsino}, we see a significant change in the result at each order in the calculation.  Furthermore, there is an additional large change still to be determined as endpoint corrections will be needed to restore a positive cross section.  We see that Sommerfeld enhancement and including large Sudakov and endpoint logarithms will all be necessary to give a controlled result.  The fact that the LL curve touches the projected CTA (the successor experiment to HESS) limits at 1 TeV only deepens the intrigue. The possibility for the next generation of experiments to probe this motivated limit hangs in the balance.
\end{enumerate}

To proceed with the analysis requires a two-step EFT. This involves matching the full theory to a SCET-I theory with a power counting parameter $\lambda = 1-z_{cut}$, followed by matching to the SCET-II EFT developed in this paper with the modified parameter $\lambda= M_W/[M_{\chi}(1-z_{cut}) ]$.  This will be the subject of a forthcoming paper.

\section*{Acknowledgements}
We thank Grigory Ovanesyan, Matthew Reece, Ira Rothstein, Tracy Slatyer, Iain Stewart and Brock Tweedie for discussions. The authors are supported by DOE grants DOE DE-FG02-04ER41338 and FG-02-06ER41449.  MB is also supported by DOE grant DE-SC0003883.

\appendix

\section{Neutralino Masses and Mixing} 

For more details on the neutralino spectrum in the limit of nearly-pure higgsino LSP, we refer the reader to the discussion in \cite{Nagata:2014wma}.  In the gauge-eigenstate basis $\chi^0$= $\left(\tilde B, \tilde W^0, \tilde h^0_d, \tilde h^0_u\right)$, the neutralino mass part of the Lagrangian is 
 \bea
 \mathcal{L} \supset -\frac{1}{2} (\chi^0)^TM_{\tilde N}\chi^0 + {\rm c.c.}, 
 \eea 
 where 
 \bea
 M_{\tilde N} =\left (\begin{array}{cccc}
 M_1 & 0 & -c_{\beta}s_Wm_Z & s_{\beta}s_Wm_Z  \\
 0 & M_2 & c_{\beta}c_W m_Z & -s_{\beta} c_W m_Z \\
 -c_{\beta}s_Wm_Z & c_{\beta}c_Wm_Z & 0 & -\mu \\
 s_{\beta}s_Wm_Z & -s_{\beta}c_Wm_Z & -\mu & 0 \end{array} \right)
 \eea
where $s_{\beta}= \sin \beta, \,\, c_W= \cos \theta_W$, etc.  Following \cite{Martin:1997ns}, we expand the eigenvalues in the limit that $m_Z \ll \mu,\, M_1,\, M_2$.  For the case of higgsino LSP, we can additionally expand in $\mu/M_{1,2}$, as we will always consider scenarios with at least a factor of a few hierarchy to avoid large mixing effects.  In this case, the neutralino mass eigenststes are very nearly a bino-like $\tilde N_1 \approx \tilde B$; a wino-like $\tilde N_2\approx \tilde W^0$;  and higgsino-like $\tilde N_3,\tilde N_4 \approx (\tilde h^0_u \pm \tilde h^0_d)/\sqrt{2}$, with mass eigenvalues:
 \bea
m_{\tilde N_1} &\approx &M_1 - \frac{m_z^2s_W^2(M_1+\mu \sin 2\beta)}{\mu^2 -M_1^2} \approx M_1+\frac{m_z^2s_W^2}{M_1}\nn\\
m_{\tilde N_2} &\approx &M_2 - \frac{m_W^2(M_2+\mu \sin 2\beta)}{\mu^2 -M_2^2} \approx M_2+\frac{m_W^2}{M_2} \nn\\
m_{\tilde N_3},m_{\tilde N_4} &=& |\mu|+\frac{m_z^2(I- \sin 2\beta)(\mu+M_1c_W^2+M_2s_W^2)}{2(\mu+M_1)(\mu+M_2)} \nn \\ 
&\approx& |\mu|+\frac{m_z^2(I-\sin 2\beta )(M_1c_W^2+M_2s_W^2)}{2M_1M_2}, \nn\\
&=& |\mu|+\frac{m_z^2(I+ \sin 2\beta)(\mu-M_1c_W^2-M_2s_W^2)}{2(\mu-M_1)(\mu-M_2)} \nn \\
&\approx& |\mu| - \frac{m_z^2(I+\sin 2\beta )(M_1c_W^2+M_2s_W^2)}{2M_1M_2},
\eea 
where $M_1$ and $M_2$ are real and positive and $\mu$ is real with $I= \pm1$.  A similar analysis for the chargino reveals mass eigenvalues for the wino-like and higgsino-like chargino
\bea
m_{\tilde C_1} =M_2 -\frac{m_W^2(M_2+\mu \sin 2\beta)}{\mu^2-M_2^2} \approx M_2 +\frac{m_W^2}{M_2} \nn\\
m_{\tilde C_2} = |\mu|+\frac{m_W^2I(\mu +M_2 \sin 2\beta)}{\mu^2-M_2^2} \approx |\mu|+\frac{m_W^2 \sin 2\beta}{M_2}  
\eea
 
For the case $I=-1$, and $M_1 \sim M_2$ we get a mass splitting between the higgsino-like neutral states to be 
\bea
\Delta M = \frac{m_Z^2}{M_1} 
\eea
This mass difference is of the order of 10-100 KeV for $M_2 \sim 10^9-10^8$  GeV. For direct detection experiments, the LSP can scatter into the heavier neutralino via a $Z$ exchange with a nucleus in the detector.  Current experiments such as XENON10 \cite{Angle:2009xb}, XENON100 \cite{Aprile:2012nq} and LUX \cite{Akerib:2013tjd} have sensitivities to $\Delta M_N \leq$ (120-200) keV in the case of higgsino DM. This sets the lower limit for the mass splitting and requires $M_1, M_2 \leq 10^5$ TeV. 

The LSP ($\tilde N_3$) and the higgsino-like chargino ($\tilde C_2$) are almost degenerate with a mass splitting 
\bea
\Delta M_+ = \frac{2M_W^2 \sin 2\beta +M_z^2(1+\sin 2\beta)}{2M_1},
\eea
which is again in the range of 100s of keV for $M_1= M_2 \approx 10^5$ TeV.
However, these are only the tree level values. 
It turns out that after including radiative corrections, $\Delta M_N$ remains of the same order, while $\Delta M_+$ is modified substantially \cite{Pierce:1996zz}. This pushes up the mass splitting to $\sim$ 350 MeV asymptotically in the large gaugino mass limit.

\section{Sommerfeld Enhancement}
\label{sec:se}

\subsection{Wino}

In order to quantify the semi-inclusive rate calculation, we need to determine the wavefunctions-at-the-origin (Sommerfeld enhancement factors) that enter our final, LL$^\prime$ cross sections in Eqs.~\ref{eq:diffrate} and \ref{crossfin}. They can be computed in principle in the nonrelativistic effective theory by summing the ladder exchange of electroweak gauge bosons between neutralinos to all orders.  Fortunately, this is equivalent to the operationally simpler task of solving the Schr\"odinger equation for our two, two-body states $| \chi^0 \chi^0 \rangle$ and $| \chi^+ \chi^- \rangle$ in the presence of the electroweak potential \cite{Hisano:2004ds,Iengo:2009ni,Cassel:2009wt}.  Since it contains Coulomb, Yukawa, and mass-shift pieces and is off-diagonal for the two states, we solve it numerically, in a manner similar to \cite{Cohen:2013ama}.  As expected for slowly moving particles in the presence of an attractive potential, we find Sommerfeld enhancement for the annihilation.  For some regions of $M_{\chi}$, this is orders of magnitude above the perturbative rate.

Taking into account appropriate state normalization, the Schr\"odinger potential is
\beq
V(r) = \left( \begin{array}{cc}
2 \, \delta M -\frac{\alpha}{r} - \alpha_W c_W^2 \frac{ e^{-m_Z r}}{r} & -\sqrt{2}\alpha_W \frac{e^{-m_W r}}{r}   \\
-\sqrt{2}\alpha_W  \frac{e^{-m_W r}}{r}  &  0 
\end{array} \right),
\label{eq:potlw}
\eeq
where $\delta M \equiv M_{\chi^+} - M_{\chi^0}$.  For numerical analysis, we use $\delta M$ = 0.17 GeV, which is its value over much of MSSM parameter space. We refer the reader to \cite{Baumgart:2014saa} for details about these enhancement factors for the wino.

\begin{figure}
\centerline{\scalebox{0.19}{\includegraphics{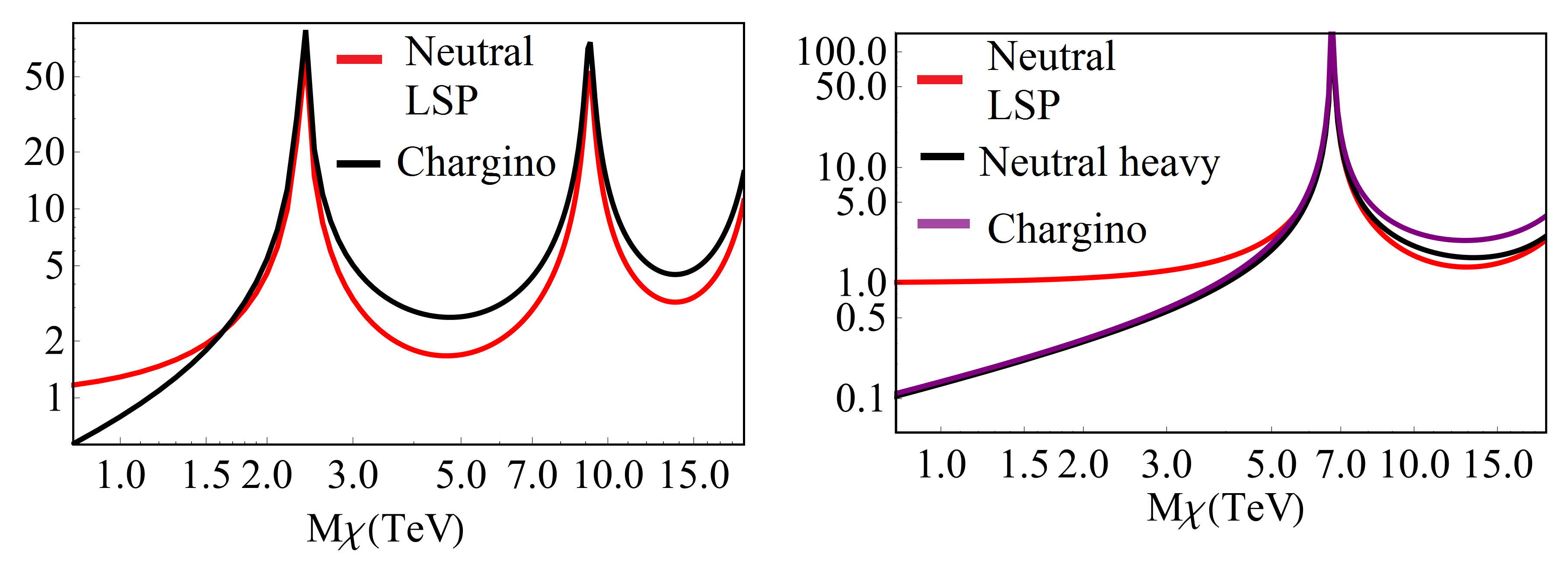}}}
\vskip-0.2cm
\caption[1]{Sommerfeld enhancement factors for wino (higgsino) in the left (right) panel. For the higgsino case, we have $\Delta M$ = 200 keV and $\Delta M_+$ = 350 MeV.}
\label{fig:somm} 
\end{figure}

\subsection{Higgsino}

For the case of the higgsino, we have three possible channels of annihilation to photons, the LSP neutralino, the heavier neutralino, and the chargino. The initial LSP state can oscillate into the heavier neutralino via $Z$ exchange or to the chargino state via the $W$ boson.  Taking into account appropriate state normalization, the Schr\"odinger potential is
\bea
V(r) = \left( \begin{array}{ccc}
2\delta m -\frac{\alpha}{r} - \frac{\alpha_W(1-2 c_W^2)^2}{4c_W^2} \frac{ e^{-m_Z r}}{r} & -\sqrt{2}\alpha_W \frac{e^{-m_W r}}{r}  & -\sqrt{2}\alpha_W \frac{e^{-m_W r}}{4r}  \\
-\sqrt{2}\alpha_W  \frac{e^{-m_W r}}{4r}  &  0  & -\frac{\alpha_W e^{-m_Z r}}{4c_W^2 r} \\
-\sqrt{2}\alpha_W  \frac{e^{-m_W r}}{4r}  &  -\frac{\alpha_W e^{-m_Z r}}{4c_W^2 r}  & 2\delta m_N \end{array} \right),
\label{eq:potlh}
\eea
We plot our numerically determined Sommerfeld factors in Fig.~\ref{fig:somm}.  These are the $\psi_{1 \textendash3}(0)$ in Eqs.~\ref{eq:wfxn} and \ref{eq:hwfxn}. Interestingly, we find that at the thermal relic mass, 1 TeV, the charged-state wavefunction-squared is $\sim 10^{-2}$.  Since charged annihilation is tree-level, we find that it is comparable to the neutral-state annihilation that occurs at one loop.  We see this reflected in the higgsino annihilation plotted in Fig.~\ref{higgsino}.

\section{Photon Wavefunction Renormalization}
\label{sec:pwr}

The wavefunction renormalization of the photon is the residue of the pole of the two-point function obtained after applying radiative corrections to $\Pi_{\gamma\gamma}$ (Fig. \ref{colld}). In the preceding calculation , we have put a generic IR cut-off at the scale $M_W$ which then gives us a correction of the form $\beta
_0 \log(\frac{2M_{\chi}}{M_W})$. While this will work for the gauge bosons and scalars which have masses around the electroweak scale, it will clearly not do for the fermionic loops . This is due to the fact that since the photon is massless, the IR cutoff for the loop integrals is the mass of the particle in the loop.  Thus, the contribution of the fermions needs to be modified to give $\sim \log(\frac{2M_{\chi}}{m_f})$.  While this is fine for the case of leptons and heavy quarks, the case of the light quarks is not so clear due to the non-perturbative physics involved. 

In principle, the way to do it formally is to match the EFT in the present form to an EFT below the electroweak scale. Such an EFT will necessarily have the SU(2) symmetry broken, with the $W$ and $Z$ bosons integrated out. The degrees of freedom that remain are the light fermions and the photon (along with the initial state winos/higgsinos). Since we are only computing diagrams up to one loop, the photon fragmentation function in this EFT will receive only a virtual correction which is the photon self energy correction  $\Pi_{\gamma \gamma}$ involving light fermionic loops. This in turn tells us that operationally the way to implement this correction in the final cross section is to identify the photon self energy terms and add $\Pi_{\gamma \gamma}$ evaluated at the scale $M_W$. Such terms are easy enough to identify since they contribute to the beta function of both the SU(2) and hypercharge gauge coupling. Thus we modify our expression for the beta function as
\bea
\frac{\alpha^i}{2\pi} \beta_0^i \log(\frac{2M_{\chi}}{M_W}) \rightarrow  \frac{\alpha^i}{2\pi} \beta_0^i \log(\frac{2M_{\chi}}{M_W}) +\Pi_{\gamma \gamma}(M_W^2),
\label{wfrnmod}
\eea  
where $\alpha^i$ can be either $\alpha_W$ or $\alpha'$ with the corresponding $\beta$ functions, $\beta_0$ or $\beta_0'$. 
We use this to modify our final LL$^\prime$ rate expressions, Eqs.~\ref{eq:diffrate} and \ref{crossfin}.  We get such a simple form in Eq.~\ref{wfrnmod} because our operators in the unbroken theory already explicitly project onto a photon external state.  Thus, they already contain the appropriate $\sin \theta_W$ factors needed for going to the broken theory below $M_W$ and no further rotation is needed.

Since $\Pi_{\gamma \gamma}$ involves non perturbative QCD corrections, we cannot compute it analytically. There is however, a method of obtaining the contribution of the quarks to the photon self-energy  \cite{Burkhardt:2001xp}: the evaluation of the loop diagrams can be related to the cross-section measurements. The imaginary part of $\Pi_{\gamma \gamma}$  for hadrons is directly related to $R_{had}$, the QED cross-section of the process $e^+ e^- \rightarrow$ hadrons normalized to the QED cross section for muon pair production. 
\bea
 Im \Pi_{\gamma \gamma}(s) = -\frac{\alpha}{3} R_{had}(s)
\eea
The real part of $\Pi_{\gamma \gamma}$ is obtained by using the Kramers-Kronig rule.  Since we have included a $\log(\frac{2M_{\chi}}{M_W})$ factor, the missing pieces can be recovered by adding the leptonic and hadronic contributions below the scale $M_W$. For the $\Pi_{\gamma \gamma}$ in Eqs.~\ref{eq:diffrate} and \ref{crossfin}, we therefore simply get $\Pi_{\gamma \gamma}$ = $\alpha(0)/\alpha(M_Z)-1$, where $\alpha$ is the fine-structure constant.  This contribution has been evaluated numerically to be -0.0594. 
\vspace{-0.2in}


\begin{thebibliography}{99}

\bibitem{Dimopoulos:1990gf} 
  S.~Dimopoulos,
  Phys.\ Lett.\ B {\bf 246}, 347 (1990).
,
\bibitem{Fan:2013faa}
J.~Fan and M.~Reece,
  JHEP {\bf 1310}, 124 (2013)
  [arXiv:1307.4400 [hep-ph]].  

\bibitem{constraint} 
  J.~Hisano, S.~Matsumoto, M.~Nagai, O.~Saito and M.~Senami,
  Phys.\ Lett.\ B {\bf 646}, 34 (2007)
  [hep-ph/0610249].
,
M.~Cirelli, A.~Strumia and M.~Tamburini,
  Nucl.\ Phys.\ B {\bf 787}, 152 (2007)
  [arXiv:0706.4071 [hep-ph]].

\bibitem{Cohen:2013ama} 
  T.~Cohen, M.~Lisanti, A.~Pierce and T.~R.~Slatyer,
  JCAP {\bf 1310}, 061 (2013)
  [arXiv:1307.4082].
  
  \bibitem{hess} 
  A.~Abramowski {\it et al.}  [HESS Collaboration],
  Phys.\ Rev.\ Lett.\  {\bf 110}, 041301 (2013)
  [arXiv:1301.1173 [astro-ph.HE]].

\bibitem{Baumgart:2014vma} 
   M.~Baumgart, I.~Z.~Rothstein and V.~Vaidya,
  arXiv:1409.4415 [hep-ph].

\bibitem{Baumgart:2014saa} 
  M.~Baumgart, I.~Z.~Rothstein and V.~Vaidya,
  JHEP {\bf 1504}, 106 (2015)
  [arXiv:1412.8698 [hep-ph]].

\bibitem{NRQCD} 
  W.~E.~Caswell and G.~P.~Lepage,
  Phys.\ Lett.\ B {\bf 167}, 437 (1986),
  \bibitem{SCET} 
  C.~W.~Bauer, S.~Fleming and M.~E.~Luke,
  Phys.\ Rev.\ D {\bf 63}, 014006 (2000)
  [hep-ph/0005275],
 C.~W.~Bauer, S.~Fleming, D.~Pirjol and I.~W.~Stewart,
  Phys.\ Rev.\ D {\bf 63}, 114020 (2001)
  [hep-ph/0011336]
  ,
  C.~W.~Bauer, D.~Pirjol and I.~W.~Stewart,
  Phys.\ Rev.\ D {\bf 65}, 054022 (2002)
  [hep-ph/0109045],
C.~W.~Bauer, S.~Fleming, D.~Pirjol, I.~Z.~Rothstein and I.~W.~Stewart,
  Phys.\ Rev.\ D {\bf 66}, 014017 (2002)
  [hep-ph/0202088].
  
  
  \bibitem{DiCintio:2013qxa} 
  A.~Di Cintio, C.~B.~Brook, A.~V.~Maccio, G.~S.~Stinson, A.~Knebe, A.~A.~Dutton and J.~Wadsley,
  Mon.\ Not.\ Roy.\ Astron.\ Soc.\  {\bf 437}, 415 (2014)
  [arXiv:1306.0898 [astro-ph.CO]].  
%
\bibitem{Kuhlen:2012qw} 
  M.~Kuhlen, J.~Guedes, A.~Pillepich, P.~Madau and L.~Mayer,
  Astrophys.\ J.\  {\bf 765}, 10 (2013)
  [arXiv:1208.4844 [astro-ph.GA]].

\bibitem{Marinacci:2013mha} 
  F.~Marinacci, R.~Pakmor and V.~Springel,
  Mon.\ Not.\ Roy.\ Astron.\ Soc.\  {\bf 437}, 1750 (2014)
  [arXiv:1305.5360 [astro-ph.CO]].

\bibitem{Bauer:2014ula} 
  M.~Bauer, T.~Cohen, R.~J.~Hill and M.~P.~Solon,
  arXiv:1409.7392 [hep-ph].
%

\bibitem{Ovanesyan:2014fwa} 
  G.~Ovanesyan, T.~R.~Slatyer and I.~W.~Stewart,
  arXiv:1409.8294 [hep-ph].

\bibitem{book_ira}
Ira Rothstein,
{\it Effective Field Theory}, $To$ $appear$

\bibitem{futuremv} 
  M.~Baumgart and V.~Vaidya,
  {\it To appear.}
  
  
  \bibitem{RRG} 
  J.~Y.~Chiu, A.~Jain, D.~Neill and I.~Z.~Rothstein,
  JHEP {\bf 1205}, 084 (2012)
  [arXiv:1202.0814 [hep-ph]]
  ,  J.~Y.~Chiu, A.~Jain, D.~Neill and I.~Z.~Rothstein,
  Phys.\ Rev.\ Lett.\  {\bf 108}, 151601 (2012)
  [arXiv:1104.0881 [hep-ph]].
,  
\bibitem{Hisano:2004ds} 
  J.~Hisano, S.~Matsumoto, M.~M.~Nojiri and O.~Saito,
  Phys.\ Rev.\ D {\bf 71}, 063528 (2005)
  [hep-ph/0412403].

\bibitem{Nagata:2014wma} 
  N.~Nagata and S.~Shirai,
  JHEP {\bf 1501}, 029 (2015)
  [arXiv:1410.4549 [hep-ph]].

 \bibitem{Bergstrom:2005ss} 
  L.~Bergstrom, T.~Bringmann, M.~Eriksson and M.~Gustafsson,
  Phys.\ Rev.\ Lett.\  {\bf 95}, 241301 (2005)
  [hep-ph/0507229].
\bibitem{Iengo:2009ni} 
  R.~Iengo,
  JHEP {\bf 0905}, 024 (2009)
  [arXiv:0902.0688 [hep-ph]].
\bibitem{Cassel:2009wt} 
  S.~Cassel,
  J.\ Phys.\ G {\bf 37}, 105009 (2010)
  [arXiv:0903.5307 [hep-ph]].
  
 \bibitem{Angle:2009xb} 
 Phys.\ Rev.\ D {\bf 80}, 115005 (2009)
 [arXiv:0910.3698 [astro-ph.CO]].
  
\bibitem{Aprile:2012nq} 
  E.~Aprile {\it et al.} [XENON100 Collaboration],
  Phys.\ Rev.\ Lett.\  {\bf 109}, 181301 (2012)
  [arXiv:1207.5988 [astro-ph.CO]].
\bibitem{Akerib:2013tjd} 
  D.~S.~Akerib {\it et al.} [LUX Collaboration],
  Phys.\ Rev.\ Lett.\  {\bf 112}, 091303 (2014)
  [arXiv:1310.8214 [astro-ph.CO]].

\bibitem{Pierce:1996zz} 
  D.~M.~Pierce, J.~A.~Bagger, K.~T.~Matchev and R.~j.~Zhang,
  Nucl.\ Phys.\ B {\bf 491}, 3 (1997)
  [hep-ph/9606211].

\bibitem{Martin:1997ns} 
  S.~P.~Martin,
  Adv.\ Ser.\ Direct.\ High Energy Phys.\  {\bf 21}, 1 (2010)
  [Adv.\ Ser.\ Direct.\ High Energy Phys.\  {\bf 18}, 1 (1998)]
  [hep-ph/9709356].
  
\bibitem{Burkhardt:2001xp} 
  H.~Burkhardt and B.~Pietrzyk,
  Phys.\ Lett.\ B {\bf 513}, 46 (2001).

\end{thebibliography}
\end{document}